\documentclass[a4paper,11pt]{article}
\usepackage{jheppub} 
\usepackage{lineno}
\usepackage{cancel}
\usepackage{abraces}
\usepackage{amsfonts,amsmath, amssymb, mathtools}
\usepackage{alphalph,etoolbox}

\patchcmd{\subequations}{\alph{equation}}{\alphalph{\value{equation}}}{}{}
\usepackage{stackengine}
\usepackage{slashed}
\usepackage{xcolor}
\usepackage{multirow}
\newcommand{\WW}{\mathcal{W}}
\newcommand{\GG}{\mathcal{G}}
\newcommand{\FF}{\mathcal{F}}
\newcommand{\AAA}{\mathcal{A}}
\newcommand{\BB}{\mathcal{B}}
\newcommand{\CC}{\mathcal{C}}
\newcommand{\HH}{\mathcal{H}}

\newcommand{\DD}{\mathcal{D}}
\newcommand{\EE}{\mathcal{E}}
\makeatletter
\def\mathcolor#1#{\@mathcolor{#1}}
\def\@mathcolor#1#2#3{%
  \protect\leavevmode
  \begingroup
    \color#1{#2}#3%
  \endgroup
}
\makeatother

\title{\boldmath $SU(2)\times SU(2)$ dilaton Weyl multiplets for maximal conformal supergravity in four, five, and six dimensions}

\author{Soumya Adhikari \& Bindusar Sahoo}
\affiliation{Indian Institute of Science Education and Research Thiruvananthapuram,\\
	Vithura, Thiruvananthapuram, India}
\emailAdd{ssoumya.a012@gmail.com, bsahoo@iisertvm.ac.in}

\abstract{New dilaton Weyl multiplets are constructed in four and five space-time dimensions for $N=4$ and $N=2$ conformal supergravity respectively. They are constructed from a mixture of the old dilaton weyl multiplets with an on-shell vector multiplet. The old dilaton Weyl multiplets have a $USp(4)$ R-symmetry group whereas the new multiplets have $SU(2)\times SU(2)$ R-symmetry, which is a subgroup of $USp(4)$. In six dimensions, for the first time, we construct a dilaton Weyl multiplet for $(2,0)$ conformal supergravity from a mixture of the standard Weyl multiplet and a tensor multiplet. The R-symmetry group for the dilaton Weyl multiplet in six dimensions is also $SU(2)\times SU(2)$.}

\begin{document}

\maketitle
\flushbottom
\vspace{-1cm}
	\section{Introduction}
Conformal supergravity, as the name suggests, is a theory of supergravity that is an extension of supergravity that enjoys only super-Poincar{\'e} symmetries, often dubbed as Poincar{\'e} supergravity \cite{Freedman:2012zz,Kaku:1978nz}. The additional local symmetries present in a theory of conformal supergravity, such as scaling or dilatations, special conformal transformations, and its superpartner special-supersymmetry, make the construction of such theories more tractable. These symmetries, together with the super-Poincar{\'e} symmetries, are known as superconformal symmetries. The construction of conformal supergravity theories often appears as an intermediate step in the construction of Poincar{\'e} supergravity theories which appears as an effective low-energy description of an ultraviolet complete theory such as String Theory. 

The entire set of procedures that allows the construction of Poincar{\'e} supergravity theories from conformal supergravity is known as the superconformal multiplet calculus \cite{VanProeyen:1983wk}. The most important ingredients in this procedure are the basic multiplets of conformal supergravity, such as the Weyl multiplet, and additional matter multiplets. The Weyl multiplet is a gauge multiplet of conformal supergravity and is known in its standard form, also known as the ``standard Weyl multiplet''  in different dimensions as well as different amount of supersymmetries \cite{Bergshoeff:1980is,Bergshoeff:1999db,Bergshoeff:1985mz,Bergshoeff:2001hc}. The standard Weyl multiplet has all the gauge fields of conformal supergravity together with some additional covariant fields. A theory of pure Poincar{\'e} supergravity, i.e., without any coupling to additional matter multiplets is gauge equivalent to a theory of conformal supergravity theory coupled to additional matter multiplets, which are also known as compensating multiplets, as these provide the necessary degrees of freedom to compensate for the extra symmetries present in conformal supergravity as compared to Poincar{\'e} supergravity. This is also known as the gauge equivalence principle \cite{Freedman:2012zz,Kaku:1978ea}. Any additional matter multiplet coupled to conformal supergravity will be physical in nature and will give rise to a matter coupled Poincar{\'e} supergravity theory.

The Weyl multiplet is an off-shell multiplet, i.e., the superconformal symmetries close on the Weyl multiplet without the necessity to impose any equations of motion. In fact, this property is one of the reasons that makes the construction of conformal supergravity theories more tractable. However, the compensating matter multiplets needed to go from conformal supergravity to Poincar{\'e} supergravity may not realize the superconformal symmetries in an off-shell manner, and one may need to impose the equations of motion for the closure of the algebra of symmetries. As a consequence of this, the pure Poincar{\'e} supergravity that one obtains by coupling compensating matter multiplets to conformal supergravity is an on-shell theory as opposed to conformal supergravity, which is off-shell. For example, $N = 4$ vector multiplets in four dimensions is an on-shell multiplet. Six of these vector multiplets coupled to conformal supergravity are required to get Poincar{\'e} supergravity \cite{deRoo:1984zyh} and hence $N=4$ Poincar{\'e} supergravity obtained in \cite{deRoo:1984zyh} is an on-shell theory.

It has been seen that one can combine a compensating matter multiplet with the standard Weyl multiplet and use the equations of motion of the matter multiplet to solve for some of the covariant fields of the standard Weyl multiplet in terms of the fields present in the matter multiplet and possibly some dual gauge fields. This combination of the compensating matter multiplet and the standard Weyl multiplet becomes a new off-shell multiplet since the equations of motion of the matter multiplet are already solved. Typically these multiplets have all the gauge fields of the standard Weyl multiplet and some of the covariant matter fields of the standard Weyl multiplet would have been replaced by the fields of the matter multiplet, including gauge fields and scalar fields (if any) and their dual. Since these multiplets will typically inherit scalar field(s) of non-vanishing Weyl weight from the matter multiplet, they are referred to as ``dilaton Weyl multiplet''. Such dilaton Weyl multiplets are, for example, constructed in the case of $N=2$ conformal supergravity in four dimensions in \cite{Butter:2017pbp} that uses a combination of vector multiplet and standard Weyl multiplet and in \cite{Gold:2022bdk} that uses a combination of hypermultiplet and standard Weyl multiplet, referred to as the hyper-dilaton Weyl multiplet. The dilaton Weyl multiplets have also been constructed for various dimensions and supersymmetries such as for $N=1$ conformal supergravity in five dimensions in \cite{Bergshoeff:2001hc,Fujita:2001kv}, $(1,0)$ conformal supergravity in six dimensions \cite{Bergshoeff:1985mz}, $N=2$ conformal supergravity in five dimensions \cite{Adhikari:2023tzi} as well as $N=4$ conformal supergravity in four dimensions \cite{Ciceri:2024xxf}. For $N=1$ conformal supergravity in five dimensions and $(1,0)$, conformal supergravity in six dimensions, hyper-dilaton Weyl multiplets have been constructed in \cite{Hutomo:2022hdi}.

Since by construction, the dilaton Weyl multiplet already has a compensating multiplet inbuilt, one would need fewer compensating multiplets to go from conformal supergravity to Poincar{\'e} supergravity. This has been seen in the construction of $N=2$ Poincar{\'e} supergravity in four dimensions \cite{Mishra:2020jlc} and $N=1$ Poincar{\'e} supergravity in five dimensions \cite{Ozkan:2013nwa} using dilaton Weyl multiplet. In both the cases a linear multiplet has been used as a compensator whereas if a standard Weyl multiplet was used, one would have needed a vector multiplet and a linear multiplet as compensating multiplets. As seen from the discussion in the previous paragraph, the dilaton Weyl multiplet is off-shell irrespective of whether the compensating matter multiplet used in its construction is off-shell or not. Hence, in the case where the compensating multiplet is on-shell, such as $N=4$ vector multiplet or $N=2$ hypermultiplet in four dimensions, one can view the dilaton Weyl multiplet as a package where one of the compensating multiplets has been upgraded from on-shell to off-shell status. As a consequence of this, one may expect fewer on-shell compensating multiplets to go from conformal to Poincar{\'e} supergravity using dilaton Weyl multiplets. One natural question to ask is, how far this program can be pushed. Can one follow a similar construction to upgrade the status of more compensating multiplets from on-shell to off-shell until one has exhausted all the compensating matter multiplets? Ultimately will it lead to the construction of a Poincar{\'e} supergravity theory which is completely off-shell that is constructed just out of the dilaton Weyl multiplet without the necessity of any compensating matter multiplet?

In order to answer the above questions, we take a very simple first step in this paper. We take the $N=4$ dilaton Weyl multiplet, which is constructed in \cite{Ciceri:2024xxf} coupled with another vector multiplet, and obtain a new dilaton Weyl multiplet by solving the equations of motion of the vector multiplet. In this way, we upgrade one more vector multiplet from on-shell to off-shell status.

We would expect this new $N=4$ dilaton Weyl multiplet in four dimensions to be related to a six-dimensional $(2,0)$ Weyl multiplet by dimensional reduction on a 2-Torus. However, it cannot be the already known $(2,0)$ standard Weyl multiplet in six dimensions \cite{Bergshoeff:1999db}, as it has already been seen in \cite{Ciceri:2024xxf} that this multiplet is related to the old $N=4$ dilaton Weyl multiplet in four dimensions by dimensional reduction along a 2-Torus. Hence, it seems plausible that there must exist a $(2,0)$ dilaton Weyl multiplet in six dimensions that can be related to the new $N=4$ dilaton Weyl multiplet in four dimensions by dimensional reduction along a 2-Torus. We construct this multiplet in six dimensions by coupling an on-shell tensor multiplet with the $(2,0)$ standard Weyl multiplet and solving the tensor multiplet equations of motion. 

As discussed in \cite{Adhikari:2023tzi}, there is no $N=2$ standard Weyl multiplet in five dimensions because of the non-existence of a rigid $N=2$ superconformal algebra in five dimensions according to a classification by Nahm \cite{Nahm:1977tg}. However, one can obtain an $N=2$ dilaton Weyl multiplet in five dimensions by reducing the six-dimensional $(2,0)$ standard Weyl multiplet along a circle. Hence, similarly, one would expect that the dimensional reduction of a six-dimensional $(2,0)$ dilaton Weyl multiplet along a circle gives rise to a new $N=2$ dilaton Weyl multiplet in five dimensions. We construct this new $N=2$ dilaton Weyl multiplet directly in five dimensions by coupling an on-shell vector multiplet to the old $N=2$ dilaton Weyl multiplet in five dimensions and solving the vector multiplet equations of motion.

To summarize, we construct new dilaton Weyl multiplets for maximal conformal supergravities in four and five dimensions, and for the first time we construct a dilaton Weyl multiplet for maximal conformal supergravity six space-time dimensions. The multiplets must be related to each other via dimensional reduction although we do not show it explicitly in this paper and leave it for future work. One common thing among all these dilaton Weyl multiplets is that their R-symmetry is $SU(2)\times SU(2)$ as opposed to the $USp(4)$ R-symmetry for the old dilaton Weyl multiplets in four and five dimensions and the standard Weyl multiplet in six dimensions. 

 The rest of the paper is organized as follows: In section-\ref{maxconf}, we discuss the existing literature on maximal conformal supergravity in four, five, and six dimensions. In sections-\ref{4dnewdilaton} and \ref{5dnewdilaton}, we give the details of the constructions of the new dilaton Weyl multiplets in four and five-dimensional maximal conformal supergravity respectively. In section-\ref{6dnewdilaton}, we give the details of the construction of a dilaton Weyl multiplet in six-dimensional maximal conformal supergravity. Finally, in section-\ref{conclusion}, we summarize and conclude with some future directions to pursue. 

\section{Maximal conformal supergravity in four, five, and six dimensions}\label{maxconf}
The maximal conformal supergravity theories in four, five, and six dimensions are referred to as $N=4$, $N=2$, and $(2,0)$, respectively. All of them have 16 ordinary or Q-supercharges and 16 special or S-supercharges, i.e., 32 supercharges in total. These supercharges are arranged in two sets of four Majorana spinors in four dimensions, two sets of $USp(4)$-symplectic Majorana spinors in five dimensions, and two sets of left chiral $USp(4)$-symplectic Majorana spinors in six dimensions. 

The key constituents in a theory of conformal supergravity are the various multiplets that form an irreducible representation of the underlying superconformal algebra. The most crucial multiplet among them is the Weyl multiplet which contains the graviton and its superpartner gravitino. The standard Weyl multiplet for $N=4$ conformal supergravity in four dimensions \cite{Bergshoeff:1980is} and $(2,0)$ conformal supergravity in six dimensions \cite{Bergshoeff:1999db} are obtained by gauging the $su(2,2|4)$ superconformal algebra and $OSp(8*|4)$ superconformal algebra, which are the superconformal algebra with 32 supercharges in four and six dimensions respectively. However, by Nahm's classification \cite{Nahm:1977tg} there exists no conformal superalgebra with 32 supercharges in five dimensions. Hence, one cannot construct any standard Weyl multiplet by gauging any superconformal algebra containing 32 supercharges in five dimensions. However, one does obtain a Weyl multiplet in five dimensions by dimensionally reducing the $(2,0)$ standard Weyl multiplet in six dimensions on a circle. This is the dilaton Weyl multiplet for $N=2$ conformal supergravity in five dimensions \cite{Adhikari:2023tzi}.

One can also construct a dilaton Weyl multiplet for $N=4$ conformal supergravity in four dimensions \cite{Ciceri:2024xxf} by coupling an on-shell $N=4$ vector multiplet to the $N=4$ standard Weyl multiplet and using the vector multiplet equations of motion to replace some of the covariant fields of the standard Weyl multiplet by some fields belonging to the vector multiplet and also a dual gauge field. This $N=4$ dilaton Weyl multiplet is in fact related to the $(2,0)$ standard Weyl multiplet in six dimensions via dimensional reduction on a 2-Torus as shown in \cite{Ciceri:2024xxf}. 

Apart from the Weyl multiplet, the other conformal supergravity multiplets that play a crucial role are the matter multiplets. For maximal conformal supergravities, these are the vector multiplets for $N=4$ and $N=2$ conformal supergravity in four \cite{Ciceri:2024xxf} and five \cite{Adhikari:2023tzi} dimensions, respectively, and the tensor multiplet for $(2,0)$ conformal supergravity in six dimensions \cite{Bergshoeff:1999db}.

The multiplets that play a crucial role in our current analysis are the $N=4$ dilaton Weyl multiplet in four dimensions, $N=2$ dilaton Weyl multiplet in five dimensions, and $(2,0)$ standard Weyl multiplet in six dimensions, respectively. In addition to the Weyl multiplets, the matter multiplets that play a crucial role in our analysis are the $N=4$ and $N=2$ vector multiplets in four and five dimensions, respectively, and the $(2,0)$ tensor multiplet in six dimensions. We give the field contents of all the relevant multiplets in the tables-\ref{dilaton4d}, \ref{4dvector}, \ref{dilaton5d}, \ref{vector5d}, \ref{table6dWeyl} and \ref{table6dtensor}. For the transformation laws, we refer the reader to the papers: \cite{Ciceri:2024xxf} for $N=4$ dilaton Weyl multiplet, \cite{Adhikari:2023tzi} for $N=2$ dilaton Weyl multiplet and vector multiplet in five dimensions, and \cite{Bergshoeff:1999db} for $(2,0)$ standard Weyl multiplet and tensor multiplet in six dimensions. The field equations for an $N=2$ vector multiplet coupled to the dilaton Weyl multiplet in five dimensions and $(2,0)$ tensor multiplet coupled to the standard Weyl multiplet in six dimensions can also be found in the above-mentioned papers.

\begin{table}[t!]
\centering
\centering
\begin{tabular}{|c|c|c|c|c|}
\hline
Field&Properties&USp(4) irreps& $w$ &$ c$\\
\hline
\multicolumn{5}{|c|}{Independent Gauge fields}\\
\hline
$e_\mu^a$&vierbein&\bf{1}&$-1$&0\\
$\psi_\mu^I$ & $\gamma_*\psi_{\mu}^{I}=\psi_{\mu}^{I}$, \text{Gravitino}&\bf{4}&$-1/2$&$-1/2$\\
$V_\mu^{IJ}$&$V_\mu^{IJ}=V_\mu^{JI}$, $USp(4)$ gauge field&\bf{10}&0&0\\
&$(V_{\mu}^{IJ})^*=\Omega_{IK}\Omega_{JL}V_{\mu}^{KL}$&&&\\
$\mathcal{A}_{\mu;\alpha}$& U(1) gauge field, $SU(1,1)$ doublet $\alpha=1,2$&\bf{1}&0&0\\
\hline
\multicolumn{5}{|c|}{Matter fields}\\
\hline
$X_a^{IJ} $ & $X_a^{IJ}=-X_a^{JI}$, $\Omega_{IJ} X_a^{IJ}=0 $& \bf{5}&1&0\\
& $(X_{a}^{IJ})^*\equiv X_{a,IJ}=-\Omega_{IK}\Omega_{JL}X_{a}^{KL}$&&&\\
$\rho$& Real Scalar&\bf{1}&1&0\\
$\phi_\alpha$& $\phi_\alpha \phi^\alpha=1, \phi^1=\phi_1^*,\phi^2=-\phi_2^*$&\bf{1}&0&$-1$\\
   $E_{IJ}$& $E_{IJ}=E_{JI}$& \bf{10}&1&$-1$\\
$T^{IJ}_{ab}$  & $T^{IJ}_{ab}=-T^{JI}_{ab}$, $T^{IJ}_{ab}=-T^{JI}_{ba}$&\bf{5}&1&$-1$\\
&$\Omega_{IJ}T^{IJ}_{ab}=0$&&&\\
$D^{IJ}{}_{KL}$& $D^{IJ}{}_{KL}=\Omega^{IM}\Omega^{JN}\Omega_{KP}\Omega_{LQ}D^{PQ}{}_{MN}$&\bf{14}&2&0\\
&$D_{KL}{}^{IJ}\equiv (D^{KL}{}_{IJ})^*=D^{IJ}{}_{KL}$&&&\\
&$D^{IJ}{}_{KJ}=0$, $\Omega_{IJ}D^{IJ}{}_{KL}=0 $&&&\\
$\chi_K{}^{IJ}$ & $\chi_K{}^{IJ}=-\chi_K{}^{JI}$,$\Omega_{IJ}\chi_K{}^{IJ}=0$&\bf{16}&$3/2$&$-1/2$\\
&$\chi_I{}^{IJ}=0$, $\gamma_*\chi_{K}{}^{IJ}=\chi_{K}{}^{IJ}$ &&&\\
$\psi_I$ & $\gamma_* \psi_I =-\psi_I$& $\bf{4}$&$3/2$&$-1/2$\\
$\Lambda_I$ & $\gamma_* \Lambda _I=\Lambda_I$ &\bf{4}& $1/2$& $-3/2$\\
\hline
\end{tabular}
\caption{Four Dimensional $N=4$ Dilaton Weyl Multiplet}
\label{dilaton4d}	
\end{table}

 \begin{table}[t!]
\centering
\centering
\begin{tabular}{ |c|c|c|c|c| }
\hline
Field & Properties & USp(4) Irreps&$w$&$c$\\
\hline
$W_\mu$& Real vector gauge field&\bf{1}&0&0\\
$\zeta_I$&Majorana, $\Gamma_*\zeta_I=-\zeta_I$&$\bf{\bar{4}}$&$3/2$&$-1/2$\\
$\phi_{IJ}$&$\phi^{IJ}\equiv(\phi_{IJ})^*=-\Omega^{IK}\Omega^{JL}\phi_{KL} $,&\bf{5}&1&0\\
    &$\Omega^{IJ}\phi_{IJ}=0$&&&\\
  $\sigma$ &Real & \bf{1}& 1& 0\\
		\hline
	\end{tabular}
	\caption{$N=4$ vector Multiplet in four dimensions}
	\label{4dvector}	
\end{table}

\begin{table}[t!]
\centering
\centering
\begin{tabular}
{|c|c|c|c|}
\hline
Field&Properties&USp(4) irreps& $w$\\
\hline
\multicolumn{4}{|c|}{Independent Gauge fields}\\
\hline
$e_\mu^a$&vielbein&\bf{1}&$-1$\\
$\psi_\mu^I$& Gravitino&\bf{4}&$-1/2$\\
$V_\mu^{IJ}$&$V_\mu^{IJ}=V_\mu^{JI}$, $USp(4)$ gauge field&\bf{10}&0\\
$C_\mu$& Graviphoton vector&\bf{1}&0\\
\hline
\multicolumn{4}{|c|}{Auxiliary matter fields}\\
\hline
$\alpha$& Real Scalar-Dilaton&\bf{1}&1\\
$E^{IJ}$ & $E^{IJ}=E^{JI}$ & \bf{10} & 1\\
$\lambda^I$&Dilatino&\bf{4}&$1/2$\\
$T^{IJ}_{ab}$  & $T^{IJ}_{ab}=-T^{JI}_{ab}$, $T^{IJ}_{ab}=-T^{IJ}_{ba}$, $\Omega_{IJ}T^{IJ}_{ab}=0$&\bf{5}&1\\
$D^{IJ,KL}$& $D^{IJ,KL}=-D^{JI,KL}=-D^{IJ,LK},$&\bf{14}&2\\
&$D^{IJ,KL}=D^{KL,IJ}, \Omega_{IJ}D^{IJ,KL}=\Omega_{IK}\Omega_{JL}D^{IJ,KL}=0.$&&\\
$\chi_K{}^{IJ}$ &$\chi_K{}^{IJ}=-\chi_K{}^{JI}, \Omega_{IJ}\chi_K{}^{IJ}=0, \chi_I{}^{IJ}=0$&\bf{16}&$3/2$\\
\hline
\end{tabular}
\caption{Five Dimensional $N=2$ Dilaton Weyl Multiplet}
		\label{dilaton5d}	
	\end{table}
	
\begin{table}[h!]
\centering
\centering
\begin{tabular}{|c|c|c|c|}
\hline
Field & Properties & USp(4) Irreps&$w$\\
\hline
$A_\mu$& Real vector&\bf{1}&0\\
$\psi^I$&Symplectic Majorana spinor&\bf4&$3/2$\\
$\sigma^{IJ}$&$\sigma^{IJ}=-\sigma^{JI},\Omega_{IJ}\sigma^{IJ}=0$&\bf5&1\\
\hline
\end{tabular}
\caption{$N=2$ vector multiplet in five dimensions}
\label{vector5d}	
\end{table}

\begin{table}[t!]
\centering
\centering
\begin{tabular}{ |c|c|c|c| }
\hline
Field&Properties&USp(4) irreps&$w$\\
\hline
\multicolumn{4}{|c|}{Independent Gauge fields}\\
\hline
$e_\mu{}^a$&vielbein &\bf{1}&-1\\
$\psi_\mu^I$&$\Gamma_*\psi_\mu^I=\psi_\mu^I$\;, gravitino &\bf{4}&$-1/2$\\
$V_\mu^{IJ}$&$V_\mu^{IJ}=V_\mu^{JI}$\;, $USp(4)$-gauge field &\bf{10}&0\\
\hline
\multicolumn{4}{|c|}{Auxiliary matter fields}\\
\hline
$T^{IJ}_{abc}$  & $\Omega_{IJ}T^{IJ}_{abc}=0, T^{IJ}_{abc}=-\frac{1}{6}\epsilon_{abcdef}T^{IJ}_{def}$&\bf{5}&1\\
$D^{IJ,KL}$& $D^{IJ,KL}=-D^{JI,KL},D^{IJ,KL}=D^{KL,IJ}$&\bf{14}&2\\
&$\Omega_{IJ}D^{IJ,KL}=0,\Omega_{IK}\Omega_{JL}D^{IJ,KL}=0$&&\\
$\chi_K{}^{IJ}$ &$\chi_K{}^{IJ}=-\chi_K{}^{JI}$, $\Omega_{IJ}\chi_K{}^{IJ}=0$&\bf{16}&$3/2$\\
&$\chi_I{}^{IJ}=0\;, \Gamma_* \chi_K{}^{IJ}=+\chi_K{}^{IJ}$&&\\
		\hline
	\end{tabular}
	\caption{Six Dimensional $(2,0)$ Standard Weyl Multiplet}
	\label{table6dWeyl}	
\end{table}

The $N=4$ vector multiplet in four dimensions coupled to a standard Weyl multiplet is given in \cite{deRoo:1984zyh}. However, we need to couple the $N=4$ vector multiplet to a dilaton Weyl multiplet for our work. In order to do that, we decompose the fields of the vector multiplet into irreducible representations of $USp(4)$ and present the transformation laws by replacing the auxiliary fields of the standard Weyl multiplet in terms of the dilaton Weyl multiplet fields already derived in \cite{Ciceri:2024xxf}.

\begin{table}[h!]
	\centering
	\centering
	\begin{tabular}{ |c|c|c|c|c|}
		\hline
		Field & Type& Properties & USp(4) Irreps&$w$\\
		\hline
		$B_{\mu \nu}$& Boson& real antisymmetric tensor gauge field,&\bf{1}&0\\
     && self-dual field strength&&\\   
		$\psi^I$&Fermion&Symplectic Majorana, $\Gamma_*\psi^I=-\psi^I$&\bf{4}&$\frac{5}{2}$\\
		$\phi^{IJ}$&Boson&$\phi^{IJ}=-\phi^{JI}, \Omega_{IJ}\phi^{IJ}=0$&\bf{5}&2\\
		\hline
	\end{tabular}
	\caption{The $(2,0)$ Tensor Multiplet in six dimensions}
	\label{table6dtensor}	
\end{table}
The Q and S transformation laws of the $N=4$ vector multiplet coupled to a dilaton Weyl multiplet are given as:
\allowdisplaybreaks
{\begin{align}\label{4dvectrans}
    \delta W_\mu &=\Phi\bigg(\bar{\epsilon}^I\gamma_\mu\zeta_I-2\bar{\epsilon}^I\psi_\mu^J\phi_{IJ}-\frac{1}{2}\bar{\epsilon}^I\psi_\mu^J\sigma\Omega_{IJ}+\bar{\epsilon}_I\gamma_\mu\Lambda_J\phi^{IJ}+\frac{1}{4}\bar{\epsilon}_I\gamma_\mu\Lambda_J\sigma\Omega^{IJ} \bigg)+\text{h.c.} \nonumber\\
    \delta\zeta_I &= -\frac{1}{2}\gamma^{ab}\tilde{\mathcal{W}}_{ab}\epsilon_I-2\cancel{D}\phi_{IJ}\epsilon^J-4\phi_{K[I}\Omega_{J]L}X_a^{KL}\epsilon^J-\sigma X_{aIJ}\epsilon^J-\frac{1}{2}\cancel{D}\sigma \epsilon^J\Omega_{IJ} \nonumber\\
    &+E_{IJ}\phi^{JK}\epsilon_K+\frac{1}{4}\sigma\Omega^{JK}E_{IJ}\epsilon_K+\frac{1}{2}\epsilon_I\bar{\Lambda}_J\psi^J-\epsilon_J\bar{\Lambda}_I\psi^J+\frac{1}{2}\gamma^a\epsilon^K\bar{\Lambda}^L\gamma_a\Lambda_I\phi_{KL} \nonumber\\
    &+\frac{1}{8}\sigma\gamma^a\epsilon^K\bar{\Lambda}^L\gamma_a\Lambda_I\Omega_{KL}-2\phi_{IJ}\eta^J-\frac{1}{2}\sigma\Omega_{IJ}\eta^J -\zeta_J k(\epsilon)^J{}_I\nonumber\\
    \delta \phi_{IJ} &=  2\bar{\epsilon}_{[I}\zeta_{J]}-\varepsilon_{IJKL}\bar{\epsilon}^K\zeta^L+2k(\epsilon)^K{}_{[I}\phi_{J]K} -\Omega\text{-trace} \nonumber\\
    \delta\sigma &=2\Omega^{IJ}\bar{\epsilon}_I\zeta_J+\Omega^{IJ}k(\epsilon)^K{}_I\phi_{JK}+\text{h.c.}\;, 
\end{align}}
where $k(\epsilon)^{I}{}_{J}=\frac{4}{\rho}\Omega^{IL}\left(\bar{\epsilon}_{[J}\psi_{L]}-\text{h.c}-\Omega\text{trace}\right)$.

The field equations of the $N=4$ vector multiplet coupled to a dilaton Weyl multiplet in four dimensions are given as:
\allowdisplaybreaks
{\begin{subequations}\label{fieldeqn4dvec}
    \begin{align}
    &D_a ( \mathcal{G}^{+ab}-\mathcal{G}^{-ab} ) = 0 \;, \label{maxeqn}
     \\[5pt]
    &\cancel{D}\zeta_I+\slashed{X}^{JK}\zeta_J\Omega_{KI}
    + \frac{1}{4}\gamma\cdot\tilde{{W}} \Lambda_I + \frac{1}{2}E_{IJ}\zeta^J -\frac{1}{4}\varepsilon_{IJKL}\gamma\cdot T^{JK}\zeta^L \nonumber\\
	&-\chi_I{}^{KL}\phi_{KL} -\frac{1}{6}\phi_{IJ}\Omega^{JK}\mathring{\chi}_K-\frac{5\sigma}{24}\mathring{\chi}_i+ \frac{1}{6}\phi_{IJ}E^{JK}\Lambda_K+ \frac{\sigma}{24}\Omega_{IJ}E^{JK}\Lambda_K\nonumber\\
 &+\frac{1}{3} \slashed{\bar{P}}\phi_{IJ}\Lambda^J+\frac{\sigma}{12} \slashed{\bar{P}}\Omega_{IJ}\Lambda^J- \frac{1}{8}\gamma_a \zeta_J \bar\Lambda_I\gamma^a\Lambda^J = 0  \;, \label{zeta} \\[5pt]
&D^aD_a\phi_{IJ}+4D^a\phi_{K[I}X_a^{KL}\Omega_{J]L}+2\phi_{K[I}\Omega_{J]L}D^aX_a^{KL} +D^a\sigma \Omega_{KI}\Omega_{JL}X_a^{KL}\nonumber\\
&+\frac{1}{2}\sigma \Omega_{KI}\Omega_{JL}D^aX_a^{KL}+\sigma\Omega_{KM}\Omega_{NI}\Omega_{JL}X^{a,MN}X_a^{KL} -2\phi_{MK}\Omega_{N[I}\Omega_{J]L}X^{a,MN}X_a^{KL}\nonumber\\
&-2\Omega_{KN}\phi_{M[I}\Omega_{J]L}X^{aMN}X_a^{KL}-T_{IJ}\cdot \tilde{F}+\Omega_{IK}\Omega_{LJ}T^{KL}\cdot \tilde{F} + \bar{\chi}^K{}_{IJ}\zeta_K +  {\frac{1}{6}}\mathring{\bar{\chi}}^{L}\Omega_{L[I}\zeta_{J]}\nonumber\\
		&+\bigg(\Omega_{IK}\Omega_{LJ}{\bar{\chi}}_M{}^{KL}-\frac{1}{6}\Omega_{M[I}\mathring{\bar{\chi}}_{J]} \bigg)\zeta^M + \frac{1}{2}\bigg( D^{KL}{}_{IJ}\phi_{KL}-\frac{2}{3}\phi_{IJ}\mathring{D}+\sigma \mathring{D}_{IJ}\bigg) - \frac{1}{6}(2\bar \zeta_{[I} \slashed{P} \Lambda_{J]}\nonumber\\
  & - \varepsilon_{IJKL} \bar \zeta^{K} \slashed{\bar{P}} \Lambda^L) - \frac{1}{12}(2\bar \Lambda^K\zeta_{[I}E_{J]K} - \varepsilon_{IJKL}\bar\Lambda_M\zeta^K E^{LM})- \frac{1}{12}\phi_{IJ}E_{KL}E^{KL}\nonumber\\
		& + \frac{1}{3}\phi_{IJ}P_{a}\bar{P}^{a} + \frac{1}{12}\phi_{IJ}(\bar \Lambda^K\slashed{D} \Lambda_K + \bar{\Lambda}_{K}\slashed{D}\Lambda^{K}) + \frac{1}{8}\phi_{IJ} \bar{\Lambda}^{K} \Lambda^{L}\bar{\Lambda}_{K} \Lambda_{L}-\Omega\text{-trace} = 0 \;, \label{phi} \\[5pt]
   & \frac{1}{\rho}D^a \{ \rho D_a \sigma -\sigma D_a\rho -4\phi_{KL}X_a^{KL}+4(\zeta^I\gamma_a\psi_I+\text{h.c.} )\} + 8\phi_{MK}\Omega_{NL}X^{a MN}X_a^{KL} \nonumber\\ 
   &=\frac{4}{\rho}(\mathcal{W}_{ab;\alpha}^+ \tilde{F}^{ab}-\mathcal{F}_{ab;\alpha}^+\tilde{W}^{ab} )\phi^\alpha +\frac{1}{\rho}\bar{\Lambda}^I \gamma \cdot \{ (\tilde{W}-\text{h.c.})\psi_I-(\tilde{F}-\text{h.c.} )\zeta_I \} +\text{h.c.}
\end{align}
\end{subequations}}
where 
\begin{align}
    \mathcal{W}_{ab}&=2e_{[a}{}^\mu e_{b]}{}^\nu \partial_\mu W_\nu -\bigg\{\Phi \bigg(\bar{\psi}^I_{[a}\gamma_{b]}\zeta_I-\bar{\psi}^I_a\psi^J_b\phi_{IJ}-\frac{1}{4}\sigma\bar{\psi}^I_a\psi^J_b\Omega_{IJ}+\bar{\psi}_{I[a}\gamma_{b]}\Lambda_J\phi^{IJ}\nonumber \\
    &+\frac{1}{4}\sigma\bar{\psi}_{I[a}\gamma_{b]}\Lambda_J\Omega^{IJ} \bigg)+ \text{h.c} \bigg\}
\end{align}
is the fully supercovariant field strength associated with the gauge field $W_{\mu}$. The field strength $\mathcal{G}_{ab}$ appearing in the Maxwell's equation is given as:
\begin{align}
\mathcal{G}_{ab}^+&=-i\frac{\varphi}{\Phi}\mathcal{W}_{ab}^++\frac{i}{2\Phi}\bar{\Lambda}^I\gamma_{ab}\zeta_I-\frac{2i}{\Phi}T_{abIJ}\phi^{IJ}-\frac{i}{2\Phi}\sigma \mathring{T}{}^+_{ab} 
\end{align}
The expression of $\mathcal{G}_{ab}^- $ directly follows from complex conjugation of $\mathcal{G}^+_{ab} $. The field strengths $\WW_{ab}$ and $\GG_{ab}$ combines as follows to form an $SU(1,1)$ doublet \cite{Ciceri:2024xxf}
\begin{align}\label{Wdoublet}
\WW_{ab;1}=\frac{1}{2}\left(i\GG_{ab}-\WW_{ab}\right)\;, \WW_{ab;2}=\frac{1}{2}\left(i\GG_{ab}+\WW_{ab}\right)
\end{align}
One can combine the above $SU(1,1)$ doublet with the coset scalars to form an $SU(1,1)$ invariant field strength $\tilde{W}_{ab}$ that appears in the equations-\ref{4dvectrans} and \ref{fieldeqn4dvec}   
\begin{align}
    \Tilde{W}_{ab}&\equiv \varepsilon^{\alpha\beta}\phi_\alpha {\WW}_{ab;\beta} =\frac{i}{2}(\phi_1-\phi_2)\mathcal{G}_{ab}-\frac{1}{2}(\phi_1+\phi_2)\mathcal{W}_{ab}
\end{align}    

We also give the expression for the dependent fields $\mathring{T}_{ab}$, $\mathring{D}_{IJ}\;, \mathring{D}\;, \mathring{\chi}^I$ as they will be needed later.
\begin{subequations}
\begin{align}
& \mathring{T}_{ab}^{-}=\frac{2}{\rho}\left[-i\Phi^*G^{-}_{ab}-\varphi^*F^{-}_{ab}+\frac{1}{2}\bar{\Lambda}_{i}\gamma_{ab}\psi^{i}\right]=\frac{4}{\rho}\left[\varepsilon^{\alpha\beta}\mathcal{F}_{ab;\alpha}^{-}\phi_{\beta}+\frac{1}{4}\bar{\Lambda}_{i}\gamma_{ab}\psi^{i}\right] \\[5pt]
    &\mathring{D}_{IJ}= {\frac{2}{\rho}D_a\rho X^{a,KL}\Omega_{IK}\Omega_{JL}+D_aX^{a,KL}\Omega_{IK}\Omega_{JL}}-2X_a^{KL}X^{a.MN}\Omega_{KM}\Omega_{IL}\Omega_{JN}\nonumber\\
    &+\frac{1}{2}\Omega_{IJ}X_a^{KL}X^{a,MN}\Omega_{KM}\Omega_{LN} +\frac{2}{\rho}(T_{IJ}\cdot \tilde{F}-\text{h.c.})-\frac{2}{\rho}\bar{\psi}_K\chi^K{}_{IJ}-\frac{1}{3\rho}\bar{\psi}_{[I}\Omega_{J]L}\mathring{\chi}^L\nonumber\\
    & {-\frac{1}{12\rho}\Omega_{IJ}\psi_K\mathring{\chi}^K -\frac{2}{\rho}\Omega_{IK}\Omega_{LJ}\bar{\psi}^M{\chi}_M{}^{KL}+\frac{1}{3\rho}\Omega_{M[I}\bar{\psi}^M\mathring{\chi}_{J]}+\frac{1}{12\rho}\Omega_{IJ}\bar{\psi}^M\mathring{\chi}_M } \nonumber\\ 
    &+\bigg( \frac{1}{3\rho}(2\bar{\psi}_{[I}\cancel{P}\Lambda_{J]}-\varepsilon_{IJKL}\bar{\psi}^K\cancel{\bar{P}}\Lambda^L)+\frac{1}{6\rho}(2\bar{\Lambda}^K\psi_{[I}E_{J]K}-\varepsilon_{IJKL}\bar{\Lambda}_M\psi^K E^{LM})\nonumber\\
    &-\Omega\;\text{-trace}\bigg)  \\[5pt]
    &\mathring{D}={-\frac{3}{5\rho}D^aD_a\rho}-\frac{3}{5}X_a^{KL}X^{a,MN}\Omega_{KM}\Omega_{LN}+\frac{1}{20}E_{KL}E^{KL}-\frac{1}{5}P_a\bar{P}^a-\frac{3}{40}\bar{\Lambda}^K\Lambda^L\bar{\Lambda}_K\Lambda_L\nonumber\\
    &+\bigg({-\frac{1}{2\rho}\bar{\psi}_K\mathring{\chi}^K}+\frac{1}{5\rho}\bar{\psi}_I\cancel{P}\Lambda_J\Omega^{IJ}+\frac{1}{10\rho}\bar{\Lambda}^K\psi_IE_{JK}\Omega^{IJ}-\frac{1}{20}\bar{\Lambda}^K(\cancel{D}\Lambda_K +\slashed{X}^{MN}\Lambda_M\Omega_{NK}) \nonumber\\
    &+\frac{3}{5\rho}\mathring{T}^+\cdot \tilde{F}+\text{h.c.} \bigg) \\[5pt]
    &\mathring{\chi}_I= {\frac{24}{5\rho}\cancel{D}\psi_I}+\frac{24}{5\rho}\slashed{X}^{JK}\psi_J\Omega_{KI}+ {\frac{6}{5\rho}\gamma\cdot \tilde{F}\Lambda_I}+\frac{12}{5\rho}E_{IJ}\psi^J-\frac{6}{5\rho}\varepsilon_{IJKL}\gamma\cdot T^{JK}\psi^L\nonumber\\
    &+\frac{1}{5}\Omega_{IJ}E^{JK}\Lambda_K +\frac{2}{5}\bar{\cancel{P}}\Omega_{IJ}\Lambda^J-\frac{3}{5\rho}\gamma_a\psi_J\bar{\Lambda}_I\gamma^a\Lambda^J  
\end{align}
\end{subequations}
\textbf{Notation and Convention:} For four-dimensional conformal supergravity, we follow the chiral notation, where the raising and lowering of $USp(4)$ R-symmetry indices are done via complex conjugation. For instance, the bosonic field $T_{ab}{}^{IJ}$ which is the anti-self dual part of a real anti-symmetric tensor field, goes to $T_{abIJ}$ under complex conjugation, which is the self-dual part of the same real anti-symmetric tensor field. Precisely $T_{abIJ}=(T_{ab}{}^{IJ})^*$.\footnote{For an arbitrary antisymmetric tensor $t_{ab}$, the self-dual (anti-self-dual) projections are defined as
$t_{ab}^{\pm} \equiv \frac{1}{2}\bigg(t_{ab} \pm \frac{1}{2}\varepsilon_{abcd}t^{cd} \bigg)$. Hence, if $t_{ab}$ is real, then the self-dual and anti-self dual parts are related by complex conjugation. We follow the convention where the Levi civita $\varepsilon_{abcd}$ is purely imaginary and defined as $\varepsilon_{0123}=i$.} Similarly, a fermionic field $\Lambda^{I}$ which is the right chiral part of a Majorana spinor, goes to $\Lambda_I$, which is the left chiral part of the same Majorana spinor, under complex conjugation. Precisely,
\begin{align}
    \Lambda^{I}=i\gamma^{0}c^{-1}\left(\Lambda_I\right)^*\;,
\end{align} 
where $c$ is the charge conjugation matrix in four dimensions. This is precisely the Majorana condition for spinors in four dimensions that relates the left and right chiral parts of a Majorana spinor via complex conjugation, similar to how the reality condition on an antisymmetric tensor field relates its self-dual and anti-self dual part via complex conjugation. The chiral notation would be followed for all kinds of fields irrespective of whether they satisfy any reality condition or not. For example, $E_{IJ}$ is a complex field in the $N=4$ Weyl multiplet. It is related to $E^{IJ}$ via complex conjugation, i.e. $E^{IJ}=(E_{IJ})^*$. 

However, we do not follow any chiral notation for five- and six-dimensional conformal supergravity. The raising and lowering of the $USp(4)$ indices is done via the $USp(4)$ invariant $\Omega^{IJ}$ and its inverse $\Omega_{IJ}$ respectively that satisfy $\Omega^{IJ}\Omega_{KJ}=\delta^{I}{}_{K}$. For instance $\Phi^{I}=\Omega^{IJ}\Phi_{J}$, and $\Phi_{I}=\Phi^{J}\Omega_{JI}$, where $\Phi^{I}$ can be any field. The symplectic Majorana conditions in five and six dimensions also relate the spinor and its complex conjugate as:
\begin{align}
\psi^{I}=\pm i\Omega^{IJ}\Gamma^{0}C^{-1}(\psi^{J})^*
\end{align}
The $+$ sign is for five dimensions, and the $-$ sign is for six dimensions, $\Gamma^{0}$ and $C$ is the $0^{th}$ component of the gamma matrix and the charge conjugation matrix, respectively which needs to be taken in the appropriate dimensions (i.e., five or six).

All the above-mentioned notation for raising and lowering the R-symmetry indices are easily generalized to fields carrying multiple indices.
\section{The new $N=4$ dilaton Weyl multiplet in four dimensions}\label{4dnewdilaton}
The new $N=4$ dilaton Weyl multiplet in four dimensions is constructed by first coupling the old $N=4$ dilaton Weyl multiplet \cite{Ciceri:2024xxf} with an $N=4$ vector multiplet as mentioned in the previous section (tables-\ref{dilaton4d} and \ref{4dvector}, eq-\ref{4dvectrans} and eq-\ref{fieldeqn4dvec}) and then using the field equations of the vector multiplet (\ref{fieldeqn4dvec}) to solve for some components of the auxiliary fields in the old dilaton Weyl multiplet in terms of the fields in the vector multiplet as well as some dual gauge fields.\footnote{The dual gauge fields appear, as we will see when some of the field equations are solved by interpreting them as the Bianchi identities of some dual gauge fields.} The resulting multiplet, which is a combination of the remaining fields of the old dilaton Weyl multiplet, vector multiplet, and the dual gauge fields, will constitute the components of the new dilaton Weyl multiplet. This multiplet will be completely off-shell, although the vector multiplet that we used was on-shell. This is because we have already solved the vector multiplet field equations.

In order to solve the vector multiplet field equations, we first need to break the R-symmetry of the old dilaton Weyl multiplet. In particular, we would need to break the $USp(4)$ R-symmetry of the old dilaton Weyl multiplet to $SU(2)\times SU(2)$. This will ensure that some components of the auxiliary fields $D^{IJ}{}_{KL}$ and $\chi^{I}{}_{JK}$ will decouple from the remaining fields in the field equations of the scalar and the fermion, respectively, which can then be used to solve these equations. As a first step, we need to decompose the $USp(4)$ irreducible representations carried by the fields of the four-dimensional $\mathcal{N}=4$ old dilaton Weyl multiplet and the vector multiplet to $SU(2)\times SU(2)$ irreducible representations. We present the decompositions in Appendix-\ref{4dirrepsbreaking} (See Table-\ref{4dbreak}).\footnote{We denote some of the decomposed fields of the old dilaton Weyl multiplet in Table-\ref{4dbreak}, which become dependent fields in the new dilaton Weyl multiplet with an inverted hat notation.} We use the following notation. The $USp(4)$ indices are denoted as $I$, which takes the values $1 \rightarrow 4$. It decomposes into two $SU(2)$ indices: $i$ taking values $1\rightarrow 2$ and $\mathcal{A}$ taking values $\bar{1}  \rightarrow \bar{2}$. $\mathcal{A}=\bar{1},\bar{2}$ corresponds to $I=3,4$. We use the following representation of the $USp(4)$ invariant $\Omega^{IJ}$:\footnote{This representation of $\Omega^{IJ}$ is different from the one used in \cite{Ciceri:2024xxf} and are related by the orthogonal transformation $\Omega\to S^{T}\Omega S$ where $S=\begin{pmatrix}0 &0 & 0 &1 \\ 1 &0&0&0\\0&0&1&0\\0&1&0&0\end{pmatrix}$. The orthogonal transformation just rearranges some of the $USp(4)$ indices.}
\begin{align}\label{omegadef}
\Omega=\begin{pmatrix}
    \varepsilon & 0\\0 & -\varepsilon
\end{pmatrix}
\end{align}
i.e. the non trivial components of $\Omega^{IJ}$ are $\Omega^{ij}=\varepsilon^{ij}$ and $\Omega^{\mathcal{A}\mathcal{B}}=-\varepsilon^{\mathcal{A}\mathcal{B}}$, where $\varepsilon^{ij}$ and $\varepsilon^{\AAA\BB}$ are the totally anti-symmetric 2-dimensional Levi Civita defined as $\varepsilon^{12}=1=\varepsilon^{\bar{1}\bar{2}}$.
The inverse $\Omega_{IJ}$ is defined via the relation:
\begin{align}
    \Omega^{IJ}\Omega_{IK}=\delta^{J}{}_{K}
\end{align}
The $SU(4)$ invariant Levi-Civita $\varepsilon_{IJKL}$ and the $USp(4)$ invariant $\Omega_{IJ}$ satisfy the following relation inside $USp(4)$:\footnote{This relation is the same as that of \cite{Ciceri:2024xxf}.}
\begin{align}
\varepsilon_{IJKL}=-\Omega_{IJ}\Omega_{KL}+\Omega_{IK}\Omega_{JL}-\Omega_{JK}\Omega_{IL}
\end{align}
In appendix-\ref{4dirrepsbreaking}, we also give the reality conditions on the decomposed fields carrying $SU(2)\times SU(2)$ irreducible representations (see eq-\ref{reality_prop}), which are induced from the reality conditions on the original fields carrying $USp(4)$ irreducible represnetations (see table-\ref{dilaton4d} and \ref{4dvector}). In order to break the $USp(4)$ to $SU(2)\times SU(2)$, the vector multiplet field $\phi^{IJ}$ plays an important role. It decomposes into a field $\phi^{i\mathcal{A}}$ and $\phi$ as shown in appendix-\ref{4dirrepsbreaking}. We break the $USp(4)$ to $SU(2)\times SU(2)$ by using the gauge fixing condition: 
\begin{align}\label{gfcond}
    \phi^{i\mathcal{A}}=0\;.
\end{align}

Q-supersymmetry will not preserve this gauge fixing condition, and hence, the unbroken Q-supersymmetry will have to be appropriately redefined as shown below by adding a field dependent $USp(4)$ transformation:
 \begin{align}\label{newsusy}
     \delta ^{\text{new}}_Q \equiv \delta_Q(\epsilon) + \delta _{USP(4)} (\Lambda^{i\mathcal{A}}= u(\epsilon)^{i\mathcal{A}} )\;, 
 \end{align}
where $u(\epsilon)^{i\mathcal{A}}$ is an $\epsilon$ and field dependent parameter given as, 
 \begin{align}\label{comp_par_usp4}
     u(\epsilon)^{i\mathcal{A}} &= \frac{1}{2i\phi}\bigg\{\varepsilon^{ij}\varepsilon^{\mathcal{A}\mathcal{B}}\bigg(\bar{\epsilon}_j\zeta_\mathcal{B}-\bar{\epsilon}_\mathcal{B}\zeta_j\bigg)-\text{h.c.} \bigg\} 
 \end{align}
 The $USp(4)$ gauge field $V_{\mu}^{IJ}$ will decompose into objects carrying $SU(2)\times SU(2)$ representations as shown in Table-\ref{4dbreak}. $V_{\mu}^{ij}$ and $V_{\mu}^{\mathcal{A}\mathcal{B}}$ will play the role of $SU(2)\times SU(2)$ gauge fields. However, $V_a^{i\mathcal{A}}$ will have to be modified by some gravitino-dependent terms to convert it into a covariant matter field. The gravitino-dependent terms are obtained as follows. Because of the modified supersymmetry transformation laws (\ref{newsusy}), which contains the field dependent $USp(4)$ parameter $u(\epsilon)^{i\mathcal{A}}$ (\ref{comp_par_usp4}), the new supersymmetry transformation of $V_a^{i\mathcal{A}}$ will contain derivative of the supersymmetry transformation parameter $\epsilon$. These are canceled by adding explicit gravitino dependent terms to $V_a^{i\mathcal{A}}$ to convert it into a covariant matter field $Y_a^{i\mathcal{A}}$ and the relation is as given below:
 \begin{align}\label{Ydef}
     Y_{a}^{i\mathcal{A}}=V_{a}^{i\mathcal{A}}-\frac{1}{2}u(\psi_a)^{i\mathcal{A}}
 \end{align}
The supercovariant derivative $D_a$ of a field appearing in the old dilaton Weyl multiplet is supercovariant w.r.t the old supersymmetry as well as $USp(4)$ R-symmetry. For the new dilaton Weyl multiplet, it will decompose into a supercovariant derivative $D_a^{\text{new}}$, which is covariant with respect to the new supersymmetry and $SU(2)\times SU(2)$ R-symmetry, along with some supercovariant terms involving $Y_a^{i\mathcal{A}}$. For instance, for $\lambda^i, \lambda^\mathcal{A}$, we have
\begin{align}\label{Dnew4d}
D_a \lambda^i&=D^{\text{new}}_a\lambda^i+Y_a^{i\mathcal{B}}\lambda^\mathcal{C}\varepsilon_{\mathcal{B}\mathcal{C}}\nonumber\\
D_a\lambda^\mathcal{A}&=D_a^{\text{new}} \lambda^\mathcal{A}-Y_a^{j\mathcal{A}}\lambda^k\varepsilon_{jk} 
\end{align}
In what follows after this, only $D_a^{\text{new}} $ will appear, and we will drop the `new' superscript for convenience.

Now, let us rewrite the field equations-\ref{fieldeqn4dvec} in terms of the $SU(2)\times SU(2)$ representations as well as the new super covariant derivative after imposing the gauge fixing condition \ref{gfcond}.

\begin{subequations}\label{4dchid}
\begin{align}
&D_a ( \mathcal{G}^{+ab}-\mathcal{G}^{-ab} ) = 0\;, \label{4dME}  \\[5pt]
&{\cancel{D}\zeta_i}+\slashed{Y}_{i\mathcal{B}}\zeta_\mathcal{C}\varepsilon^{\mathcal{BC}}-(i\check{\slashed{X}}\zeta_i+\slashed{X}^{k\mathcal{A}}\zeta_\mathcal{A}\varepsilon_{ki} ){+\frac{1}{4}\gamma\cdot \tilde{\mathcal{W}}\Lambda_i}+\frac{1}{2}(E_{ij}\zeta^j+E_{i\mathcal{A}}\zeta^\mathcal{A})\nonumber\\
&-\frac{1}{2}(\varepsilon_{ij}\varepsilon_{\mathcal{BC}}\gamma\cdot T^{j\mathcal{B}}\zeta^\mathcal{C}+\varepsilon_{il}\gamma\cdot \check{T}^-\zeta^l) -4i\check{\chi}_i\phi+\frac{i\phi}{6}\mathring{\chi}_i-\frac{5\sigma}{24}\mathring{\chi}_i\nonumber \\
&+\frac{i}{6}\phi\varepsilon_{ij}(E^{jk}\Lambda_k+E^{j\mathcal{B}}\Lambda_\mathcal{B})+\frac{\sigma}{24}\varepsilon_{ij}(E^{jk}\Lambda_k+E^{j\mathcal{A}}\Lambda_\mathcal{A}) +\frac{i}{3}\cancel{\bar{P}}\phi\Lambda^j\varepsilon_{ij}\nonumber \\
&+\frac{\sigma}{12}\bar{\cancel{P}}\Lambda^j\varepsilon_{ij} -\frac{1}{8}\gamma_a(\zeta_j\bar{\Lambda}_i\gamma^a\Lambda^j+\zeta_\mathcal{A}\bar{\Lambda}_i\gamma^a\Lambda^\mathcal{A})=0\;, \label{zetai} \\[5pt]
&{\cancel{D}\zeta_\mathcal{A}}+\slashed{Y}_{j\mathcal{A}}\zeta_k\varepsilon^{jk}+\slashed{X}^{j\mathcal{B}}\zeta_j\varepsilon_{\mathcal{BA}}-i\check{\slashed{X}}\zeta_\mathcal{A}{+\frac{1}{4}\gamma\cdot \tilde{\mathcal{W}}\Lambda_\mathcal{A}}+\frac{1}{2}E_{\mathcal{AB}}\zeta^\mathcal{B}+\frac{1}{2}E_{j\mathcal{A}}\zeta^j \nonumber\\
&-\frac{1}{2}\varepsilon_{\mathcal{AC}}(\gamma\cdot \check{T}^-\zeta^\mathcal{C}-\varepsilon_{jk}\gamma\cdot T^{j\mathcal{C}}\zeta^k)-4i\phi\check{\chi}_\mathcal{A}-\frac{i\phi}{6}\mathring{\chi}_\mathcal{A}-\frac{5\sigma}{24}\mathring{\chi}_\mathcal{A} \nonumber \\
& +\frac{i}{6}\phi\varepsilon_{\mathcal{AB}}(E^{k\mathcal{B}}\Lambda_k+E^{\mathcal{BC}}\Lambda_\mathcal{C})-\frac{\sigma}{24}\varepsilon_{\mathcal{AB}}(E^{k\mathcal{B}}\Lambda_k+E^{\mathcal{BC}}\Lambda_\mathcal{C})+\frac{i}{3}\cancel{\bar{P}}\phi\varepsilon_{\mathcal{AB}}\Lambda^\mathcal{B}\nonumber \\
&-\frac{\sigma}{12}\varepsilon_{\mathcal{AB}}\bar{\cancel{P}}\Lambda^\mathcal{B} -\frac{1}{8}\gamma_a(\zeta_j\bar{\Lambda}_\mathcal{A}\gamma^a\Lambda^j+\zeta_\mathcal{B}\bar{\Lambda}_\mathcal{A}\gamma^a\Lambda^\mathcal{B})=0\;, \label{zetaA} \\[5pt]
&D^a D_a\phi +2\phi Y_a^{k\mathcal{C}}Y^a_{k\mathcal{C}}+4\phi X^a_{i\AAA}X_a^{i\AAA}-8\phi \check{X}^a\check{X}_a 
+i\check{T}\cdot \tilde{F}\nonumber\\
&-\check{X}^aD_a\sigma-\frac{1}{2}\sigma D^a\check{X}_a+i(\check{\bar{\chi}}^k\zeta_k-\check{\bar{\chi}}_m\zeta^m)-i(\check{\bar{\chi}}^\mathcal{A}\zeta_\mathcal{A}-\check{\bar{\chi}}_\mathcal{A}\zeta^\mathcal{A})\nonumber\\
&+2\phi \check{D}-\frac{1}{3}\phi \mathring{D}-\frac{i\sigma}{4}\varepsilon^{ij}\mathring{D}_{ij}-\frac{\phi}{12}(E_{kl}E^{kl}+E_{\mathcal{AB}}E^{\mathcal{AB}}+2E_{k\mathcal{A}}E^{k\mathcal{A}} ) +\frac{1}{3}\phi P_a\bar{P}^a \nonumber\\
&+\frac{\phi}{12}\{ \bar{\Lambda}^k(\cancel{D}\Lambda_k+\slashed{Y}^{l\mathcal{C}}\Lambda_\mathcal{C}\varepsilon_{lk} )   +\bar{\Lambda}^\mathcal{C}(\cancel{D}\Lambda_\mathcal{C}+\slashed{Y}^{l\mathcal{B}}\Lambda_l\varepsilon_{\mathcal{BC}} )+\bar{\Lambda}_k(\cancel{D}\Lambda^k+\slashed{Y}^{k\mathcal{B}}\Lambda^\mathcal{C}\varepsilon_{\mathcal{BC}} ) \nonumber\\
&+\bar{\Lambda}_\mathcal{C}(\cancel{D}\Lambda^\mathcal{C}+\slashed{Y}^{j\mathcal{C}}\Lambda^k\varepsilon_{kj} ) \} +\frac{1}{16}\phi (\bar{\Lambda}^k\Lambda^l\bar{\Lambda}_k\Lambda_l+\bar{\Lambda}^\mathcal{B}\Lambda^l\bar{\Lambda}_\mathcal{B}\Lambda_l+\bar{\Lambda}^k\Lambda^\mathcal{C}\bar{\Lambda}_k\Lambda_\mathcal{C}+\bar{\Lambda}^\mathcal{B}\Lambda^\mathcal{C}\bar{\Lambda}_\mathcal{B}\Lambda_\mathcal{C} )\nonumber\\
&+\frac{i}{24}\{\bar{\mathring{\chi}}^j\zeta_j-\bar{\mathring{\chi}}^\BB\zeta_\BB +(\bar{\Lambda}^k\zeta_iE_{jk}+\bar{\Lambda}^\BB\zeta_iE_{j\BB} )\varepsilon^{ij}+(\bar{\Lambda}^k\zeta_\AAA E_{k\BB}+\bar{\Lambda}^\mathcal{C}\zeta_\AAA E_{\BB\mathcal{C}} )\varepsilon^{\AAA\BB} \nonumber\\
&+2(\bar{\zeta}_i\cancel{P}\Lambda_j\varepsilon^{ij}+\bar{\zeta}_\AAA \cancel{P}\Lambda_\BB \varepsilon^{\AAA\BB}) -\text{h.c.}\}   =0\;,  \label{phieq} \\[5pt]
&-2i\phi D_aY^a_{i\mathcal{A}}-2i\phi(D_aX^a_{i\mathcal{A}}+2iY^a_{i\mathcal{A}}\check{X}^a )+4\phi \check{X}_aX^a_{i\mathcal{A}}+D_a\sigma X^a_{i\AAA}+\frac{\sigma}{2}D_a \check{X}^a_{i\AAA}+i\sigma Y^a_{i\AAA}\check{X}_a \nonumber\\
&+4\phi X^a_{i\mathcal{A}}\check{X}_a-T_{i\mathcal{A}}\cdot \tilde{F}-\varepsilon_{ik}\varepsilon_{\mathcal{BA}}T^{k\mathcal{B}}\cdot \tilde{F}+\frac{1}{12}(\bar{\mathring{\chi}}^l\varepsilon_{li}\zeta_\mathcal{A}+\bar{\mathring{\chi}}^\mathcal{B}\varepsilon_{\mathcal{BA}}\zeta_i)+\varepsilon^{kl}\bar{\chi}_{il;\mathcal{A}}\zeta_k  +\frac{1}{2}\varepsilon_{\mathcal{AB}}\check{\bar{\chi}}^\mathcal{B}\zeta_i\nonumber\\
&-\varepsilon^{\mathcal{BC}}\bar{\chi}_{i;\mathcal{AC}}\zeta_\mathcal{B}-\frac{1}{2}\varepsilon_{ij}\check{\bar{\chi}}^j\zeta_\mathcal{A} +\frac{1}{12}\bar{\mathring{\chi}}^l\varepsilon_{li}\zeta_\mathcal{A}-\frac{1}{12}\bar{\mathring{\chi}}^\mathcal{B}\zeta_i\varepsilon_{\mathcal{AB}}+\varepsilon_{ik}\varepsilon_{\mathcal{AB}}\varepsilon_{mn}\bar{\chi}^{nk;\mathcal{B}}\zeta^m -\frac{1}{2}\varepsilon_{im}\check{\bar{\chi}}_\mathcal{A}\zeta^m \nonumber\\
& -\varepsilon_{ik}\varepsilon_{\mathcal{AB}}\varepsilon_{\mathcal{CD}}\bar{\chi}^{k;\mathcal{BD}}\zeta^\mathcal{C}+\frac{1}{2}\varepsilon_{\mathcal{AC}}\check{\bar{\chi}}_i\zeta^\mathcal{C}-\frac{1}{12}\varepsilon_{mi}\bar{\mathring{\chi}}_\mathcal{A}\zeta^m-\frac{1}{12}\varepsilon_{\mathcal{BA}}\bar{\mathring{\chi}}_i\zeta^\mathcal{B}+2i\phi \check{D}_{i\mathcal{A}}+\frac{1}{2}\sigma \mathring{D}_{i\mathcal{A}}\nonumber\\
&-\frac{1}{6}\{\bar{\zeta}_i\cancel{P}\Lambda_\mathcal{A}-\bar{\zeta}_\mathcal{A}\cancel{P}\Lambda_i-\varepsilon_{ik}\varepsilon_{\mathcal{AB}}(\bar{\zeta}^\mathcal{B}\bar{\cancel{P}}\Lambda^k-\bar{\Lambda}^k\bar{\cancel{P}}\Lambda^\mathcal{B} )\} -\frac{1}{12}\bigg\{\bar{\Lambda}^k\zeta_iE_{k\mathcal{A}}+\bar{\Lambda}^\mathcal{B}\zeta_iE_{\mathcal{AB}} \nonumber\\
&-\bar{\Lambda}^k\zeta_\mathcal{A}E_{ik}-\bar{\Lambda}^\mathcal{B}\zeta_\mathcal{A}E_{i\mathcal{B}}-\varepsilon_{il}\varepsilon_{\mathcal{AB}}(\bar{\Lambda}_m\zeta^\mathcal{B}E^{lm}-\bar{\Lambda}_m\zeta^lE^{m\mathcal{B}}+\bar{\Lambda}_\mathcal{C}\zeta^\mathcal{B}E^{l\mathcal{C}}-\bar{\Lambda}_\mathcal{C}\zeta^lE^{\mathcal{BC}} ) \bigg\} =0\;, \label{phiiA}  \\[5pt]
&D^a \{\rho D_a\sigma -\sigma D_a\rho +16\rho \phi \check{X}_a+4(\bar{\zeta}^i\gamma_a\psi_i+\bar{\zeta}^\mathcal{A}\gamma_a\psi_\mathcal{A}+\text{h.c.}) \} \nonumber  \\
   &=-\frac{4}{\rho}(\mathcal{W}_{ab;\alpha}^+ \tilde{F}^{ab}-\mathcal{F}_{ab;\alpha}^+\tilde{W}^{ab} )\phi^\alpha +\frac{1}{\rho}\bar{\Lambda}^i \gamma_{ab} \{ (\tilde{W}^{ab}-\text{h.c.})\psi_i-(\tilde{F}^{ab} -\text{h.c.} )\zeta_i \}\nonumber\\
   & +\frac{1}{\rho}\bar{\Lambda}^\mathcal{A} \gamma_{ab} \{ (\tilde{W}^{ab}-\text{h.c.})\psi_\mathcal{A} -(\tilde{F}^{ab} -\text{h.c.} )\zeta_\mathcal{A} \}+\text{h.c.}\;. \label{sigmaeq}
\end{align}
\end{subequations}

The field equation for $\zeta^I$ (\ref{zeta}) decomposes into the field equations for $\zeta^i$ and $\zeta^\mathcal{A}$. They are given in (\ref{zetai}) and (\ref{zetaA}). It can be seen that the fields $\chi^{ij;\mathcal{A}} $ and $\chi^{i;\mathcal{AB}}$ drop out from the fields equations as a result of the gauge condition (\ref{gfcond}) and the fields $\check{\chi}^i$, $\check{\chi}^\mathcal{A}$ can now be solved algebraically using these field equations and are completely determined in terms of the other fields.  The fields $\chi^{ij;\mathcal{A}} $ and $\chi^{i;\mathcal{AB}}$ remain as independent fields.

Similarly, field equations for $\phi_{IJ} $ (\ref{phi}) decomposes into the field equations for $\phi$ and $\phi^{i\mathcal{A}}$. They are given in (\ref{phieq}) and (\ref{phiiA}). It can be seen that $D^{ij;\mathcal{AB}}$ drops out from the field equations as a result of the gauge condition (\ref{gfcond}) and the fields $\check{D}$, $\check{D}^{i\mathcal{A}}$ can now be solved algebraically using these field equations. The field $D^{ij;\AAA\BB}$ remains as an independent field.

The Maxwell's equation (\ref{4dME}) is solved by interpreting it as the Bianchi identity of a supercovariant field strength $\mathcal{G}_{ab}$ corresponding to a new gauge field $\tilde{W}_{\mu}$. One can then use it to solve for $\check{T}_{ab}^{\pm}$ as shown below 

\begin{align}\label{Tab}
     \check{T}^+_{ab}&=\frac{1}{4i\phi\rho}\bigg\{\rho \mathcal{W}_{ab;\alpha}\phi^\alpha-\sigma \mathcal{F}_{ab;\alpha}\phi^\alpha +\frac{1}{4}\bigg(\rho (\Lambda^i\gamma_{ab}\zeta_i+\bar{\zeta}^\mathcal{A}\gamma_{ab}\zeta_\mathcal{A})-\sigma(\bar{\Lambda}^i\gamma\psi_i+\bar{\Lambda}^\mathcal{A}\gamma_{ab}\psi_{\mathcal{A}} )\bigg) \bigg\}  \nonumber\\
     \check{T}^{-}_{ab}&=(\check{T}^+_{ab})^*
\end{align}
where $\FF_{ab;\alpha}$ is the $SU(1,1)$ doublet of field strength corresponding to the $SU(1,1)$ doublet of gauge fields $\AAA_{\mu;\alpha}$ belonging to the old dilaton Weyl multiplet. The $SU(1,1)$ doublet of field strengths $\WW_{ab;\alpha}$ is defined in (\ref{Wdoublet}) where $\GG_{\mu\nu}$ is now promoted to the status of a field strength corresponding to an off-shell gauge field $\tilde{W}_{\mu}$. This allows us to define a new $SU(1,1)$ doublet of gauge fields $\WW_{\mu;\alpha}$ in terms of the gauge fields $W_{\mu}$ and $\tilde{W}_{\mu}$, which becomes a part of the new dilaton Weyl multiplet:
\begin{align}
    \WW_{\mu;1}\equiv -\frac{1}{2}\left(W_{\mu}-i\tilde{W}_{\mu}\right)\;, \WW_{\mu;2}\equiv -\left(\WW_{\mu;1}\right)^*
\end{align}

In  order to solve the field equation for the scalar field $\sigma$ (\ref{sigmaeq}), we re-write it as follows: 
\allowdisplaybreaks{\begin{align}\label{sigmaeqH}
    D_{[a}\mathcal{H}_{bcd]}&=24i\mathcal{W}_{[ab;\alpha}\mathcal{F}_{cd];\beta}\phi^{[\alpha}\varepsilon^{\beta]\gamma}\phi_{\gamma}\nonumber\\
    &-3i(\bar{\Lambda}^i\gamma_{[ab}\psi_i+\bar{\Lambda}^\mathcal{A}\gamma_{[ab}\psi_\mathcal{A}+\text{h.c.})(\mathcal{W}_{cd];\alpha}\phi^\alpha-\mathcal{W}_{cd];\beta}\phi_\alpha\varepsilon^{\alpha\beta})\nonumber\\
&+3i(\bar{\Lambda}^i\gamma_{[ab}\zeta_i+\bar{\Lambda}^\mathcal{A}\gamma_{[ab}\zeta_\mathcal{A}+\text{h.c.})(\mathcal{F}_{cd];\alpha}\phi^\alpha-\mathcal{F}_{cd];\beta}\phi_\alpha\varepsilon^{\alpha\beta} )  \;. 
\end{align}
where the 3-form $\HH_{abc}$ is defined as the dual of a 1-form, as shown below 
\begin{align}\label{3-formdef}
   & -\frac{i}{3!}\varepsilon_a{}^{bcd}\mathcal{H}_{bcd}=\rho D_a\sigma -\sigma D_a\rho +16\rho \phi \check{X}_a+4(\bar{\zeta}^i\gamma_a\psi_i+\bar{\zeta}^\mathcal{A}\gamma_a\psi_\mathcal{A}+\text{h.c.})
\end{align}
We solve the $\sigma$-field equations the same way as the Maxwell's equation. We interpret it as the Bianchi identity of a 3-form field strength corresponding to a 2-form gauge field $B_{\mu\nu}$, which now becomes a part of the new dilaton Weyl multiplet. The equation-\ref{3-formdef} is used to solve for the auxiliary field $\check{X}_a$ in terms of the other independent fields of the multiplet. 

From the Bianchi identity (\ref{sigmaeqH}), we obtain the bosonic gauge transformations of the two-form $B_{\mu\nu}$: 
\begin{align}
    \delta B_{\mu\nu}&=2\partial_{[\mu}\Lambda_{\nu]} -8i\left(\Lambda^W_\alpha F_{\mu\nu;\beta}+\Lambda^A_\beta W_{\mu\nu;\alpha}\right)\phi^{[\alpha}\varepsilon^{\beta]\gamma}\phi_\gamma\nonumber\\
&+2i\left(\bar{\Lambda}^i\gamma_{\mu\nu}\psi_i+\bar{\Lambda}^\mathcal{A}\gamma_{\mu\nu}\psi_\mathcal{A}+\text{h.c.}\right)\left(\Lambda_\alpha^W\phi^\alpha-\Lambda_\beta^W\phi_\alpha\varepsilon^{\alpha\beta} \right) \nonumber\\
&-2i\left(\bar{\Lambda}^i\gamma_{\mu\nu}\zeta_i+\bar{\Lambda}^\mathcal{A}\gamma_{\mu\nu}\zeta_\mathcal{A}+\text{h.c.}\right)\left(\Lambda_\alpha^A\phi^\alpha-\Lambda_\beta^A\phi_\alpha\varepsilon^{\alpha\beta}\right )
\end{align}

where $\Lambda_\nu$ is the gauge transformation parameter associated with the two-form gauge field $B_{\mu\nu}$ and $\Lambda^W_\alpha\;, \Lambda^A_\alpha$ are the gauge transformation parameters associated with the $U(1)$ gauge fields $\mathcal{W}_{\mu,\alpha}$ and $\mathcal{A}_{\mu,\alpha}$ respectively. The supersymmetry transformation of the two-form $B_{\mu\nu}$ is obtained as follows. From equation-\ref{3-formdef} , we obtain the supersymmetry transformation of the 3-form field strength $H_{\mu\nu\rho}$. Then the supersymmetry transformation of $B_{\mu\nu}$ is obtained by demanding consistency with the supersymmetry transformation of the 3-form field strength and the Bianchi identity-\ref{sigmaeqH}. The result is given in \ref{two-form}. 

Now, we summarise the results. The field equations \ref{4dchid} make some of the auxiliary fields of the old dilaton Weyl multiplet composite. The remaining auxiliary fields together with the fields of the vector multiplet and some dual gauge fields combine with the independent gauge fields of the old dilaton Weyl multiplet to constitute the new dilaton Weyl multiplet of $N=4$ conformal supergravity where the R-symmetry is $SU(2)\times SU(2)$.

In Table-\ref{newdilaton4d}, we give the full details of the new dilaton Weyl multiplet.
\begin{table}[t!]
		\centering
		\centering
		\begin{tabular}{|c|c|c|c| c|}
			\hline
			Field&Properties&$SU(2)\times SU(2)$&$ w$ & $c$\\
			\hline
			\multicolumn{5}{|c|}{Independent Gauge fields}\\
			\hline
			$e_\mu^a$&vielbein&\bf{(1,1)}&$-1$&0\\
			$\psi_\mu^i $& $\gamma_*\psi^i_\mu=\psi^i_\mu$, Gravitino&$\bf{(2,1) }$&$-1/2$&$-1/2$\\
            $ \psi_\mu^{\mathcal{A}} $&  $\gamma_*\psi^i_\mu=\psi^i_\mu$, Gravitino&$\bf{ (1,2)}$&$-1/2$&$-1/2$\\
			$V_\mu^{ij}$&$V_\mu^{ij}=V_\mu^{ji}$, $SU(2)$ gauge field&\bf{(3,1)}&0&0\\
   $V_\mu^{\mathcal{A} \mathcal{B}}$&$V_\mu^{\mathcal{A} \mathcal{B} }=V_\mu^{\mathcal{B} \mathcal{A} }$, $SU(2)$ gauge field&\bf{(1,3)}&0&0\\
   $A_{\mu;\alpha}$& U(1) gauge field, $SU(1,1)$ doublet $\alpha=1,2$&\bf{(1,1)}&0&0\\
         $W_{\mu;\alpha}$& U(1) gauge field, $SU(1,1)$ doublet $\alpha=1,2$&\bf{(1,1)}&0&0\\
                  $B_{\mu\nu}$& Tensor gauge field&\bf{(1,1)}&0&0\\
			\hline
			\multicolumn{5}{|c|}{Covariant fields}\\
			\hline
 $Y_a^{i\AAA}$& Boson& \bf{2,2}&$1$&0\\           
   $X_a^{i\mathcal{A}} $ & $X_a^{IJ}=-X_a^{JI}, \Omega X_a^{IJ}=0 $& \bf{(2,2)}&1&0\\
   $\rho$& Real Scalar&\bf{(1,1)
   }&1&0\\
    $\sigma$& Real Scalar&\bf{(1,1)
   }&1&0\\
      $\phi$& Real Scalar&\bf{(1,1)}&1&0\\
   $\phi_\alpha$& $\phi_\alpha \phi^\alpha=1, \phi^1=\phi_1^*,\phi^2=-\phi_2^*$&\bf{(1,1)}&0&$-1$\\
   $E_{ij}$& $E_{ij}=E_{ji} $& (\bf{3,1})&1&$-1$\\
   $E_{\mathcal{A}\mathcal{B}}$& $E_{\mathcal{B}\mathcal{A}}=E_{\mathcal{B}\mathcal{A}} $& (\bf{1,3})&1&$-1$\\
   $E_{i\mathcal{A}}$& Bosonic field & (\bf{2,2})&$1$&$-1$\\
			$T^{i\mathcal{A}}_{ab}$  & Anti-self dual&\bf{(2,2)}&$1$&$-1$\\
     $\psi_i$ & $\gamma_* \psi_i =-\psi_i$& $\bf{(2,1)}$&$3/2$&$-1/2$\\
$   \psi_{\mathcal{A}}$  & $\gamma_* \psi_{\mathcal{A}} =-\psi_{\mathcal{A}} $& $\bf{(1,2)}$&$3/2$&$-1/2$\\
$\Lambda_i$ & $\gamma_* \Lambda _i=\Lambda_*$ &\bf{(2,1)}& $1/2$& $-3/2$\\
$\Lambda_{\mathcal{A}} $ & $\gamma_* \Lambda _{\mathcal{A}} =\Lambda_{\mathcal{A}} $ &\bf{(1,2)}& $1/2$& $-3/2$\\
$\zeta_i$ & $\gamma_* \lambda_i =-\lambda_i$& $\bf{(2,1)}$&$3/2$&$-1/2$\\
$   \zeta_{\mathcal{A}}$  & $\gamma_* \lambda_{\mathcal{A}} =-\lambda_{\mathcal{A}} $& $\bf{(1,2)}$&$3/2$&$-1/2$\\
			$\chi^{ij;\mathcal{A}} $ & $\gamma_*\chi^{ij;\AAA}=\chi^{ij;\AAA}$, $\chi^{ij;\mathcal{A}}=\chi^{ji;\mathcal{A}} $&\bf{(3,2)}&$3/2$&$-1/2$\\
   $\chi^{i;\mathcal{A}\mathcal{B}} $&  $\gamma_*\chi^{i;\mathcal{A}\mathcal{B}}=\chi^{i;\mathcal{A}\mathcal{B}}$, $\chi^{\mathcal{A}\mathcal{B};i}=\chi^{\mathcal{B}\mathcal{A };i}$& $\bf{(2,3)}$&$3/2$&$-1/2$\\
   $D^{ij;\mathcal{A}\mathcal{B}}$& $D^{ij;\mathcal{A}\mathcal{B}}=D^{ji;\mathcal{A}\mathcal{B},KL}=D^{ij,\mathcal{B}\mathcal{A}}$&\bf{(3,3)}&2&0\\
			\hline
		\end{tabular}
		\caption{Four-Dimensional New Dilaton Weyl Multiplet}
		\label{newdilaton4d}	
	\end{table} 
 In \ref{4dndiltrans}, we give the Q and S transformations of the independent gauge fields of the new dilaton Weyl multiplet. 
 \allowdisplaybreaks
{ \begin{subequations}\label{4dndiltrans}
\begin{align}
     \delta e_\mu^a&= \bar{\epsilon}^i \gamma^a\psi_{\mu i}+\bar{\epsilon}^{\mathcal{A}}\gamma^a\psi_{\mu \mathcal{A}} + \text{h.c.}\\[5pt]
     \delta \psi_\mu^i&= 2\mathcal{D}_\mu \epsilon^i +2i\check{X}_\mu\epsilon^i+2Y_\mu^{i\mathcal{A}}\varepsilon_{\mathcal{A}\mathcal{B}}\epsilon^{\mathcal{B}}+2X_\mu^{i\mathcal{A}}\varepsilon_{\mathcal{A}\mathcal{B}}\epsilon^{\mathcal{B}}-\frac{1}{8}\gamma^{ab}\left(4 \check{T}_{ab}^-+ \mathring{T}^-_{ab} \right) \gamma_\mu \epsilon_j\varepsilon^{ij} \nonumber\\
     &-\frac{1}{2}\gamma^{ab}T_{ab}^{i\mathcal{A}}\gamma_\mu\epsilon_{\mathcal{A}}+\epsilon^{ij}\epsilon^{\mathcal{A}\mathcal{B}}\left(\bar{\psi}_{\mu j}\epsilon_{\mathcal{A}}\Lambda_{\mathcal{B}}+\bar{\psi}_{\mu \mathcal{A}}\epsilon_{\mathcal{B}}\Lambda_j-\bar{\psi}_{\mu \mathcal{A}}\epsilon_j\Lambda_{\mathcal{B}}\right)+k(\epsilon)^i{}_j \psi_\mu^j\nonumber\\
&+k(\epsilon)^i{}_{\mathcal{A}}\psi_\mu^{\mathcal{A}}-k(\psi_\mu)^i{}_j\epsilon^j-k(\psi_\mu)^i{}_{\mathcal{A}}\epsilon^{\mathcal{A}}-u(\epsilon)^{i\mathcal{A}}\varepsilon_{\mathcal{A}\mathcal{B}}\psi_\mu^{\mathcal{B}}+u(\psi_\mu)^{i\mathcal{A}}\varepsilon_{\mathcal{A}\mathcal{B}} \epsilon^{\mathcal{B}}-\gamma_\mu\eta^i \\[5pt]
     \delta \psi_\mu^\mathcal{A}&=2D_\mu\epsilon^\mathcal{A}-2Y_\mu^{i\mathcal{A}}\varepsilon_{ij}\epsilon^j-2i\check{X}_\mu \epsilon^\mathcal{A}+2{X}_\mu^{j\mathcal{A}}\varepsilon_{jk}\varepsilon^k+\frac{1}{2}\gamma\cdot T^{j\mathcal{A}}\gamma_\mu\epsilon_j\nonumber\\
     &-\frac{1}{8}\gamma\cdot\left(4\check{T}^--\mathring{T}{}^-\right)\gamma_\mu\epsilon_\mathcal{B}\varepsilon^{\mathcal{A}\mathcal{B}}+k(\epsilon)^\mathcal{A}{}_j\psi^j_\mu+k(\epsilon)^\mathcal{A}{}_\mathcal{B}\psi^\mathcal{B}_\mu-k(\psi_\mu)^\mathcal{A}{}_j\varepsilon^j-k(\psi_\mu)^\mathcal{A}{}_\mathcal{B}\epsilon^\mathcal{B}\nonumber\\
     &+u(\epsilon)^{j\mathcal{A}}\psi_\mu^k\varepsilon_{jk} -u(\psi_\mu)^{j\mathcal{A}}\varepsilon_{jk}\epsilon^k+\varepsilon^{\mathcal{A}\mathcal{B}}\varepsilon^{jk}\left(\bar{\psi}_{\mu j}\epsilon_k\Lambda_\mathcal{B}-\bar{\psi}_{\mu j}\epsilon_\mathcal{B}\Lambda_k+\bar{\psi}_{\mu \mathcal{B}}\epsilon_j\Lambda_k \right)-\gamma_\mu\eta^\AAA\\[5pt]
     \delta V_\mu^{ij}&= \bar{\epsilon}^{(i}\phi_{\mu k}\varepsilon^{j)k}+\bar{\epsilon}^{(i}\gamma_{\mu}\check{\chi}^{j)}-\bar{\epsilon}^\mathcal{A}\gamma_{\mu}\chi_{kl;\mathcal{A}}\varepsilon^{ik}\varepsilon^{jl} -\frac{1}{2}\varepsilon_{kl}\varepsilon_{\mathcal{A}\mathcal{B}}\left(- \varepsilon^{k(i}E^{j)\mathcal{A}}\bar{\epsilon}^\mathcal{B}\psi_\mu^l \right.\nonumber\\
&\left.+E^{l(i}\varepsilon^{j)k}\bar{\epsilon}^\mathcal{A}\psi_\mu^\mathcal{B}+\varepsilon^{k(i}E^{j)\mathcal{A}}\bar{\epsilon}^l\psi_\mu^\mathcal{B} \right)-\frac{1}{6}E^{l(i}\bar{\epsilon}_k\gamma_\mu \Lambda_l\varepsilon^{j)k}+\frac{1}{6}\bar{\epsilon}_k\gamma_\mu \Lambda_\mathcal{B}\varepsilon^{k(i}E^{j)\mathcal{B}}\nonumber\\
&+\frac{1}{3}\bar{\epsilon}^{(i}\gamma_\mu \cancel{P}\Lambda_k\varepsilon^{j)k} +\frac{1}{4}\varepsilon^{k(i}\varepsilon^{j)l}\varepsilon^{\mathcal{A}\mathcal{B}}\left(\bar{\epsilon}_l\gamma \cdot T_{k\mathcal{A}}\gamma_\mu\Lambda_\mathcal{B}+\bar{\epsilon}_\mathcal{A}\gamma \cdot T_{k\mathcal{B}} \gamma_\mu\Lambda_l \right) \nonumber\\
&+\frac{1}{4}\varepsilon^{s(i}\left(-\bar{\epsilon}^{j)}\gamma_a\psi_{\mu s} \bar{\Lambda}_\mathcal{A}\gamma^a \Lambda^\mathcal{A}+\bar{\epsilon}^\mathcal{A}\gamma_a\psi_{\mu s}\bar{\Lambda}_\mathcal{A}\gamma^a\Lambda^{j)} +\bar{\epsilon}^{j)}\gamma_a\psi_{\mu \mathcal{A}}\bar{\Lambda}_s\gamma^a\Lambda^\mathcal{A}\right. \nonumber\\
&\left.-\bar{\epsilon}^\mathcal{A}\gamma_a\psi_{\mu \mathcal{A}}\bar{\Lambda}_s \gamma^a\Lambda^{j)}\right)-2u(\epsilon)^{(i\mathcal{A}}Y_\mu^{j)\mathcal{B}}\varepsilon_{\mathcal{A}\mathcal{B}}-u(\epsilon)^{(i\mathcal{A}}u(\psi_\mu)^{j)\mathcal{B}}\varepsilon_{\mathcal{A}\mathcal{B}}-\bar{\psi}_\mu^{(i}\eta_k\varepsilon^{j)k} +\text{h.c.}\\[5pt]
\delta V_\mu^{\mathcal{A}\mathcal{B}}&=-\bar{\epsilon}^{(\mathcal{A}}\phi_{\mu \mathcal{C}}\varepsilon^{\mathcal{B})\mathcal{C}}-\bar{\epsilon}^j\gamma_\mu \varepsilon^{\mathcal{D}(\mathcal{A}}\varepsilon^{\mathcal{B})\mathcal{C}}\chi_{j;\mathcal{C}\mathcal{D}}-\bar{\epsilon}^{(\mathcal{A}}\gamma_\mu\check{\chi}^{\mathcal{B})}+\frac{1}{2}\varepsilon_{\mathcal{C}\mathcal{D}}\varepsilon_{ls}\left( E^{l(\mathcal{A}}\varepsilon^{\mathcal{B})\mathcal{C}}\bar{\epsilon}^s\psi_\mu^\mathcal{D} \right.\nonumber\\
&\left.+E^{\mathcal{D}(\mathcal{A}}\varepsilon^{\mathcal{B})\mathcal{C}}\bar{\epsilon}^l\psi_{\mu }^s+E^{s(\mathcal{A}}\varepsilon^{\mathcal{B})\mathcal{C}}\bar{\epsilon}^\DD\psi_\mu^l \right)+\frac{1}{6}E^{l(\mathcal{A}}\bar{\epsilon}_\mathcal{C}\gamma_\mu \Lambda_l\varepsilon^{\mathcal{B})\mathcal{C}}+\frac{1}{6}E^{\mathcal{D}(\mathcal{A}}\bar{\epsilon}_\mathcal{C}\gamma_\mu\Lambda_\mathcal{D}\varepsilon^{\mathcal{B})\mathcal{C}}\nonumber\\
&-\frac{1}{3}\bar{\epsilon}^{(\mathcal{A}}\gamma_\mu \cancel{P}\Lambda_\mathcal{C}\varepsilon^{\mathcal{B})\mathcal{C}}\nonumber+\frac{1}{4}\varepsilon^{\mathcal{C}(\mathcal{A}}\varepsilon^{\mathcal{B})\mathcal{D}}\varepsilon^{sl}\left(\bar{\epsilon}_s\gamma\cdot T_{l\mathcal{C}}\gamma_\mu\Lambda_\mathcal{D}-\bar{\epsilon}_\mathcal{D}\gamma\cdot T_{l\mathcal{C}}\gamma_\mu\Lambda_s \right)\\
&-\frac{1}{4}\varepsilon^{\mathcal{C}(\mathcal{A}}\left(-\bar{\epsilon}^s\gamma_a\psi_{\mu s} \bar{\Lambda
}_\mathcal{C}\gamma^{a}\Lambda^{\mathcal{B})}+\bar{\epsilon}^{\mathcal{B})}\gamma_a\psi_{\mu s} \bar{\Lambda}_\mathcal{C}\gamma^{a}\Lambda^s+\bar{\epsilon}^s\gamma_a\psi_{\mu \CC}\bar{\Lambda}_s\gamma^a\Lambda^{\BB)}\right.\nonumber\\
&\left.-\bar{\epsilon}^{\BB)}\gamma_a\psi_{\mu \CC}\bar{\Lambda}_s\gamma^a\Lambda^s \right)+2\varepsilon_{kl}u(\epsilon)^{k(\mathcal{A}}\left(Y_\mu^{l\mathcal{B})}+\frac{1}{2}u(\psi_\mu)^{l\mathcal{B})} \right)+\bar{\psi}_\mu^{(\AAA}\eta_\mathcal{C}\varepsilon^{\BB)\mathcal{C}} +\text{h.c.} \\[5pt]
\delta W_{\mu;\alpha}&=-\phi_\alpha\left(\bar{\epsilon}_i\gamma_\mu\zeta^i+\bar{\epsilon}_\mathcal{A}\gamma_\mu\zeta^\mathcal{A}-\frac{1}{2}\left(4i\phi+\sigma\right)\bar{\epsilon}_i\psi_{\mu j}\varepsilon^{ij}+\frac{1}{2}\left(-4i\phi+\sigma\right)\bar{\epsilon}_\mathcal{A}\psi_{\mu\mathcal{B}}\varepsilon^{\mathcal{A}\mathcal{B}}\right.\nonumber\\
&\left.+\frac{1}{4}\left(4i\phi+\sigma\right)\bar{\epsilon}^i\gamma_\mu\Lambda^j\varepsilon_{ij}+\frac{1}{4}\left(4i\phi-\sigma\right)\bar{\epsilon}^\mathcal{A}\gamma_\mu\Lambda^\mathcal{B}\varepsilon_{\mathcal{A}\mathcal{B}}\right)\nonumber\\
&+\varepsilon_{\beta\alpha}\phi^\beta \bigg(\bar{\epsilon}^i\gamma_\mu\zeta_i+\bar{\epsilon}^\mathcal{A}\gamma_\mu\zeta_\mathcal{A}-\frac{1}{2}(-4i\phi+\sigma)\bar{\epsilon}^i\psi_\mu^j\varepsilon_{ij}+\frac{1}{2}(4i\phi+\sigma)\bar{\epsilon}^\mathcal{A}\psi_\mu^\mathcal{B}\varepsilon_{\mathcal{A}\mathcal{B}}\nonumber\\
&+\frac{1}{4}(-4i\phi+\sigma)\bar{\epsilon}_i\gamma_\mu\Lambda_j\varepsilon^{ij}+\frac{1}{4}(-4i\phi+\sigma)\bar{\epsilon}_\mathcal{A}\gamma_\mu\Lambda_\mathcal{B}\varepsilon^{\mathcal{A}\mathcal{B}}\bigg)\\[5pt]
\delta A_{\mu;\alpha}&=-\phi_\alpha\bigg(\bar{\epsilon}_i\gamma_\mu\psi^i+\bar{\epsilon}_\mathcal{A}\gamma_\mu\psi^\mathcal{A}-\frac{1}{2}\rho\bar{\epsilon}_i\psi_{\mu j}\varepsilon^{ij}+\frac{1}{2}\rho\bar{\epsilon}_\mathcal{A}\psi_{\mu \mathcal{B}}\varepsilon^{\mathcal{A}\mathcal{B}}+\frac{1}{4}\rho\bar{\epsilon}^i\gamma_\mu\Lambda^j\varepsilon_{ij}\nonumber\\
&-\frac{1}{4}\rho\bar{\epsilon}^\mathcal{A}\gamma_\mu\Lambda^\mathcal{B}\varepsilon_{\mathcal{AB}}\bigg)-\varepsilon_{\alpha\beta}\phi^\beta\bigg( \bar{\epsilon}^i\gamma_\mu\psi_i+\bar{\epsilon}^\mathcal{A}\gamma_\mu\psi_\mathcal{A}-\frac{1}{2}\rho\bar{\epsilon}^i\psi_\mu^j\varepsilon_{ij}+\frac{1}{2}\rho\bar{\epsilon}^\mathcal{A}\psi_\mu^{\mathcal{B}}\varepsilon_{\mathcal{A}\mathcal{B}}\nonumber\\
&+\frac{1}{4}\rho\bar{\epsilon}_i\gamma_\mu\Lambda_j\varepsilon^{ij}-\frac{1}{4}\rho\bar{\epsilon}_\mathcal{A}\gamma_\mu\Lambda_\mathcal{B}\varepsilon^{\mathcal{A}\mathcal{B}}\bigg)\\[5pt]
\delta B_{\mu\nu}&=-4i\rho\left(\bar{\epsilon}_i\gamma_{\mu\nu}\zeta_j\varepsilon^{ij}-\bar{\epsilon}_{\mathcal{A}} \gamma_{\mu\nu}\zeta_\mathcal{B}\varepsilon^{\mathcal{AB}}\right)+4i\sigma \left(\bar{\epsilon}_i\gamma_{\mu\nu}\psi_j\varepsilon^{ij}-\bar{\epsilon}_\AAA\gamma_{\mu\nu}\psi_\BB\varepsilon^{\AAA\BB}  \right)\nonumber\\
&+16\phi\left(\bar{\epsilon}_i\gamma_{\mu\nu}\psi_j\varepsilon^{ij}+\bar{\epsilon}_\mathcal{A}\gamma_{\mu\nu}\psi_\mathcal{B}\varepsilon^{\mathcal{AB}}\right)-16\phi \rho \bar{\epsilon}_i\gamma_{[\mu}\psi_{\nu]}^i+16\phi \rho \bar{\epsilon}_\mathcal{A}\gamma_{[\mu}\psi_{\nu]}^\mathcal{A} \nonumber\\
&+16i\left(W_{[\mu;\alpha}\delta_Q A_{\nu];\beta}+A_{[\mu;\beta}\delta_Q W_{\nu];\alpha} \right) \phi^{[\alpha}\varepsilon^{\beta]\gamma}\phi_\gamma+\text{h.c.} \;\label{two-form}
\end{align}
\end{subequations}
 In \ref{4dndiltranscov}, we give the Q and S transformations of the covariant fields of the new dilaton Weyl multiplet. 
\begin{subequations}\label{4dndiltranscov}

\end{subequations}}
\section{The new $N=2$ dilaton Weyl multiplet in five dimensions}\label{5dnewdilaton}
In this section, we discuss the construction of a new dilaton Weyl multiplet in $N=2$ conformal supergravity in five dimensions. We couple a vector multiplet to the old $N=2$ dilaton Weyl multiplet \cite{Adhikari:2023tzi} and use the field equations of the vector multiplet to solve for some components of the auxiliary fields of the old dilaton Weyl multiplet. Similar to the case of four dimensions discussed in the previous section, we first need to break the $USp(4)$ R-symmetry of the old dilaton Weyl multiplet to $SU(2)\times SU(2)$ in order to solve the field equations. We present the decompositions of the fields of the old dilaton Weyl multiplet and the vector multiplet in appendix \ref{5dirrepsbreaking}. We use the same notations for decomposing the $USp(4)$ indices to the $SU(2)\times SU(2)$ indices as in the case of four dimensions. The field $\sigma^{IJ} $ of the vector multiplet which is in the $\bf{5}$ of USP(4) decomposes into $\sigma^{i\mathcal{A}}$ in the $\bf{(2,2)}$ and $\sigma $ in the  $\bf{(1,1)}$ of $SU(2)\times SU(2)$ as shown in appendix \ref{5dirrepsbreaking}. We break the $USP(4)$ to $SU(2)\times SU(2)$ by using the gauge fixing condition:
\begin{align}\label{5dggcond}
 \sigma^{i\mathcal{A}}=0   
\end{align}
 This gauge fixing condition will not be preserved by Q-supersymmetry, and hence the Q-supersymmetry transformation will have to be appropriately redefined as shown below:
\begin{align}\label{compensator}
\delta_Q(\epsilon)=\delta_Q^{\text{old}}(\epsilon)+\delta_{USP(4)}(\Lambda^{i\mathcal{A}}=v(\epsilon)^{i\mathcal{A}})
	\end{align}
	
where $v(\epsilon)^{i\mathcal{A}} $ is an $\epsilon$ and field dependent $USp(4)$ parameter which is obtained by demanding that $\delta_Q\sigma^{i\mathcal{A}}=0$ and is given as:
\begin{align}
v(\epsilon)^{i\mathcal{A}}=-\frac{i}{2\sigma}\bigg(\bar{\epsilon}^{i}\psi^{\mathcal{A}}-\bar{\epsilon}^\mathcal{A}\psi^i\bigg)
\end{align}
The decomposition of the $USp(4)$ gauge field $V_{\mu}^{IJ}$ into $SU(2)\times SU(2)$ gauge fields $V_{\mu}^{ij}$ and $V_{\mu}^{\AAA\BB}$ and a covariant matter field $Y_{a}^{i\AAA}$ is essentially similar to that in four dimensions (see the discussions below eq-\ref{comp_par_usp4} till eq-\ref{Ydef}), where the latter is given as\footnote{The difference in the factor appearing before $u(\psi_{\mu})$ in \ref{Ydef} for four dimensions and before $v(\psi_\mu)$ in \ref{Y5def} below is essentially due to the different conventions of supersymmetry transformation of the gravitino in 4d and 5d}:
\begin{align}\label{Y5def}
    Y_a^{i\AAA}&=V_a^{i\AAA}-v(\psi_a)^{i\AAA}
\end{align}
As in the case of four dimensions, the old supercovariant derivative $D_a$ will also decompose into a new supercovariant derivative $D_a^{\text{new}}$, which is covariant with respect to the new supersymmetry and $SU(2)\times SU(2)$ R-symmetry, along with some supercovariant terms involving $Y_a^{i\mathcal{A}}$. For instance, for $\lambda^i, \lambda^\mathcal{A}$, we have\footnote{The difference in the factor appearing before $Y_{a}^{i\AAA}$ in eq-\ref{Dnew4d} in four dimensions and eq-\ref{Dnew5d} below is due to the difference in conventions of R-symmetry transformation in 4d and 5d.}
\begin{align}\label{Dnew5d}
D_a \lambda^i&=D^{\text{new}}_a\lambda^i+\frac{1}{2}Y_a^{i\mathcal{A}}\lambda_\mathcal{A} \nonumber\\
D_a\lambda^\mathcal{A}&=D_a^{\text{new}} \lambda^\mathcal{A}+\frac{1}{2}Y_a^{j\mathcal{A}}\lambda_j
\end{align}
In what follows after this, only $D_a^{\text{new}} $ will appear, and we will drop the `new' superscript for convenience. Now, we rewrite the field equations of the vector multiplet in $SU(2) \times 
SU(2)$ representations in equation \ref{5dfeqn} below:
\allowdisplaybreaks{ 
\begin{subequations}\label{5dfeqn}
    \begin{align}
    &D_a \mathcal{H}_{ab}=(^*\omega)_b\;,\label{5dMax}  \\[5pt]
 &\cancel{D}\psi^i+\frac{1}{2}\slashed{Y}^{i\AAA}\psi_\AAA+\frac{1}{2\alpha}\cancel{D}\alpha\psi^i+\cancel{D}\sigma\psi^i+\frac{1}{2}\sigma \slashed{Y}^{i\AAA}\lambda_\AAA+\frac{\sigma}{5\alpha}\cancel{D}\alpha\lambda^i+\frac{4}{5}\sigma\cancel{D}\lambda^i+\frac{2}{5}\sigma\slashed{Y}^{i\AAA}\lambda_\AAA \nonumber\\ 
&+\frac{i}{8\alpha}\gamma\cdot G\psi^i+\frac{i}{5\alpha}\sigma\gamma\cdot G\lambda^i+\frac{i}{2}(E^i{}_j\psi^j +E^i{}_\AAA\psi^\AAA)+\frac{i}{2}\gamma\cdot H\lambda^i+\frac{i}{10}\gamma\cdot\check{T}(2\sigma\lambda^i+5\psi^i) \nonumber\\
&+\frac{i}{10}\gamma\cdot T^{j\mathcal{A}}(2\sigma\lambda_\mathcal{A}+5\psi_\AAA)-iE^{i\AAA}\lambda_\AAA+\sigma\check{\chi}^i-\bar{\lambda}^{[i}\psi^{j]}\lambda_j-\frac{1}{2}(\bar{\lambda}^i\psi^\AAA-\bar{\lambda}^\AAA\psi^i )\lambda_\AAA\nonumber\\
&-\frac{1}{4}\lambda^i(\bar{\lambda}^j\psi_j+\bar{\lambda}^\AAA\psi_\AAA )+\frac{1}{8}(\bar{\lambda}^j\gamma_{ab}\psi_j+\bar{\lambda}^\AAA\gamma_{ab}\psi_\AAA )\gamma^{ab}\lambda^i=0 \;,\label{5dpsii}  \\[5pt]
&\cancel{D}\psi^\AAA+\frac{1}{2}\slashed{Y}^{j\AAA}\psi_j+\frac{1}{2\alpha}\cancel{D}\alpha\psi^\AAA+\cancel{D}\sigma \lambda^\AAA-\frac{\sigma}{2}\slashed{Y}^{i\AAA}\lambda_i+\frac{\sigma}{5\alpha}\cancel{D}\alpha\lambda^\AAA-\frac{4\sigma}{5}\cancel{D}\lambda^\AAA-\frac{2\sigma}{5}\slashed{Y}^{j\AAA}\lambda_j \nonumber\\
&+\frac{i}{8\alpha}\gamma\cdot G\psi^\AAA+\frac{i}{5\alpha}\sigma\gamma\cdot\lambda^\AAA-\frac{1}{2}E^{i\AAA}\lambda_i-\frac{1}{2}E^{\AAA\BB}\psi_\BB+\frac{i}{2}\gamma\cdot H \lambda^\AAA-\frac{i}{10}\gamma\cdot \check{T}(2\sigma\lambda^\AAA+5\psi^\AAA) \nonumber\\
& -\frac{i}{10} \gamma\cdot T^{k\AAA}(2\sigma\lambda_k+5\psi_k) -iE^{\AAA\BB}\lambda_\BB+\sigma\check{\chi}^\AAA+\frac{1}{2}(\bar{\lambda}^\AAA\psi^j-\bar{\lambda}^j\psi^\AAA )\lambda_j+\bar{\lambda}^{[\AAA}\psi^{\BB]}\lambda_\BB\nonumber\\
&-\frac{1}{4}(\bar{\lambda}^j\psi_j+\bar{\lambda}^\BB\psi_\BB )\lambda^\AAA+\frac{1}{8}(\bar{\lambda}^i\gamma_{ab}\psi_i+\bar{\lambda}^\BB\gamma_{ab}\psi_\BB )\gamma^{ab}\lambda^\AAA=0  \;,\label{5dpsiA} \\[5pt]
&D^aD_a\sigma-\frac{\phi}{2}Y_a^{j\BB}Y^a_{j\BB} +\frac{1}{\alpha}D^a\alpha D_a\sigma -\frac{\sigma}{5\alpha^2}D^a\alpha D_a\alpha+\frac{3\sigma}{5\alpha} D^aD_a\alpha-\frac{i}{4}E^{i\AAA}(\bar{\lambda}_\AAA\psi_i-\bar{\lambda}_i\psi_\AAA) \nonumber\\
&+\frac{i}{8}(\bar{\lambda}^i\gamma\cdot \check{T}\psi_i+\bar{\lambda}_\AAA\gamma\cdot T^{i\AAA}\psi_i )-\frac{1}{2}\bar{\lambda}^i\bigg( \frac{i}{8\alpha}\gamma\cdot G\psi_i-\frac{1}{2\alpha}\cancel{D}\alpha\psi_i+\frac{i}{2}(E_{ik}\psi^k+E_{i\BB}\psi^\BB)\nonumber\\
&+\frac{i}{2}\gamma\cdot H\lambda_i+\frac{1}{8}(\bar{\lambda}^k\gamma_{ab}\psi_k\gamma^{ab}\lambda_i-\bar{\lambda}^\BB\gamma\psi_\BB\gamma^{ab}\lambda_i+\cancel{D}\sigma\lambda_i +\frac{\sigma}{5\alpha}\cancel{D}\alpha\lambda_i+\frac{4\sigma}{5}\cancel{D}\lambda_i-\frac{2\sigma}{5}\slashed{Y}_{i\AAA}\lambda^\AAA\nonumber\\
&-\frac{i\sigma}{5\alpha}\gamma\cdot G \lambda_i-\frac{i\sigma}{5}\gamma\cdot\check {T}\lambda_i+\frac{i\sigma}{5}\gamma\cdot T_{i\AAA}\lambda^\AAA \bigg)-\frac{i\sigma}{2}\bar{\lambda}^iE_{i\AAA}\lambda^\AAA-\frac{1}{4}\bar{\lambda}^i\lambda^\AAA\bar{\lambda}_\AAA\psi_i\nonumber\\
&+\frac{1}{10}\bar{\psi}^j\bigg( \frac{1}{\alpha}\cancel{D}\alpha\lambda_j-\cancel{D}\lambda_j-\frac{i}{4\alpha}\gamma\cdot G\lambda_j+\frac{i}{4}\gamma\cdot \check{T}\lambda_j-\frac{i}{4}\gamma\cdot T_{j\AAA}\lambda^\AAA+\frac{1}{2}\slashed{Y}_{j\AAA}\lambda^\AAA\bigg)-\frac{4\sigma}{15}\check{D}\nonumber\\
&+6H\cdot \check{T}-\frac{3i}{2}\check{T}_{ab}(\bar{\lambda}^k\gamma^{ab}\psi_k+\bar{\lambda}^\BB\gamma^{ab}\psi_\BB)+\frac{i}{4}(\check{\chi}^k\psi_k+\check{\chi}^\AAA\psi_\AAA)-\frac{\sigma}{10\alpha^2}G\cdot G-\frac{2\sigma}{5}\check{T}\cdot \check{T}\nonumber\\
&-\frac{\sigma}{5}T^{i\AAA}\cdot T_{i\AAA}=0  \;,\label{5dsigma} \\[5pt]
& -\sigma D^aY_a^{i\AAA}-2Y_a^{i\AAA}D^a\sigma -\frac{\sigma}{\alpha}D^a\alpha Y_a^{i\AAA}-\frac{i}{2}E^{i\mathcal{C}}\varepsilon^{\AAA\BB}\bigg( \bar{\lambda}_{[\mathcal{C}}\psi_{\BB]}-\frac{1}{4}\varepsilon_{\mathcal{CB}}(\bar{\lambda}^p\psi_p-\lambda^\mathcal{D}\psi_\mathcal{D})\bigg) \nonumber\\
& +\frac{i}{2}E^{k\AAA}\varepsilon^{il}\bigg\{ \bar{\lambda}_{[k}\psi_{l]}+\frac{1}{4}\varepsilon_{kl}(\bar{\lambda}^p\psi_p-\lambda^\mathcal{D}\psi_\mathcal{D} )\bigg\}-\frac{1}{2}E^{\AAA\BB}\varepsilon^{il}\bigg\{\sigma E_{l\BB}-\frac{i}{2}(\bar{\lambda}_\BB\psi_l-\bar{\lambda}_l\psi_\BB ) \bigg\} \nonumber\\
&-\frac{i}{8}\left( \bar{\lambda}^i\gamma\cdot \check{T}\psi^\AAA+\bar{\lambda}_\BB\gamma\cdot T^{i\BB}\psi^\AAA+\bar{\lambda}_k\gamma\cdot T^{k\AAA}\psi^i+\bar{\lambda}^\AAA\gamma\cdot \check{T}\psi^i\right)-\frac{1}{2}\lambda^i\bigg\{ -\frac{i}{8\alpha}\gamma\cdot G\psi^\AAA \nonumber\\
&+\frac{1}{2\alpha}\cancel{D}\alpha\psi^\AAA+\frac{i}{2}(E^{k\AAA}\psi_k+E^{\AAA\BB}\psi_\BB)-\frac{i}{2}\gamma\cdot H\lambda^\AAA-\frac{1}{8}(\bar{\lambda}^k\gamma_{ab}\psi_k+\bar{\lambda}^\BB\gamma_{ab}\psi_\BB )\gamma^{ab}\lambda^\AAA\nonumber\\
&+\cancel{D}\sigma \lambda^\AAA+\frac{\sigma}{2}\slashed{Y}^{k\AAA}\lambda_k+\frac{\sigma}{5\alpha}\cancel{D}\alpha \lambda^\AAA +\frac{2\sigma}{5}(2\cancel{D}\lambda^\AAA+\slashed{Y}^{j\AAA}\lambda_j )-\frac{i\sigma}{5\alpha}\gamma\cdot G\lambda^\AAA\nonumber\\
&+\frac{i\sigma}{5}(\gamma\cdot \check{T}\lambda^\AAA-\gamma\cdot T^{j\AAA}\lambda_j) \bigg\}+\frac{1}{2}\lambda^\AAA \bigg\{ -\frac{i}{8\alpha}\gamma\cdot G\psi^i+\frac{1}{2\alpha}\cancel{D}\alpha \psi^i+\frac{i}{2}(E^{ik}\psi_k+E^{i\BB}\psi_\BB)\nonumber\\
&-\frac{i}{2}\gamma\cdot H\lambda^i-\frac{1}{8}(\bar{\lambda}^k\gamma_{ab}\psi_k+\bar{\lambda}^\BB\gamma_{ab}\psi_\BB )\gamma^{ab}\lambda^i-\cancel{D}\sigma\lambda^i-\frac{\sigma}{2}\slashed{Y}^{i\BB}\lambda_\BB-\frac{\sigma}{5\alpha}\cancel{D}\alpha\lambda^i-\frac{2\sigma}{5}(2\cancel{D}\lambda^i \nonumber\\
&+\slashed{Y}^{i\BB}\lambda_\BB )+\frac{i\sigma}{5\alpha}\gamma\cdot G\lambda^i+\frac{i\sigma}{5}(\gamma\cdot \check{T}\lambda^i+\gamma\cdot T^{i\BB}\lambda_\BB )\bigg\}-\frac{1}{2}\bar{\lambda}^i\bigg\{-i\sigma E^{j\AAA}\lambda_j+\frac{1}{2}(\bar{\lambda}^\AAA\psi^j \nonumber\\
&-\bar{\lambda}^j\psi^\AAA )\lambda_j-\bar{\lambda}^{[\AAA}\psi^{\BB]}\lambda_\BB+\frac{1}{4}\lambda^\AAA(\bar{\lambda}^j\psi_j+\bar{\lambda}^\BB\psi_\BB)\bigg\}-\frac{1}{2}\bar{\lambda}^\AAA\bigg(-i\sigma E^{i\BB}\lambda_\BB+\lambda_j\bar{\lambda}^{[i}\psi^{j]}\nonumber\\
&+\frac{1}{4}\lambda^i(\bar{\lambda}^j\psi_j+\bar{\lambda}^\BB\psi_\BB)+\frac{1}{2}(\bar{\lambda}^i\psi^\BB-\bar{\lambda}^\BB\psi^i )\lambda_\BB\bigg)+\frac{1}{10}\psi^i\bigg(\frac{1}{\alpha}\cancel{D}\alpha\lambda^\AAA-\frac{1}{2}(2\cancel{D}\lambda^\AAA+\slashed{Y}^{j\AAA}\lambda_j )\nonumber\\
&-\frac{i}{4\alpha}\gamma\cdot G\lambda^\AAA-\frac{i}{4}(\gamma\cdot T^{k\AAA}\lambda_k+\gamma\cdot \check{T}\lambda^\AAA)\bigg)-\frac{1}{10}\psi^\AAA\bigg(\frac{1}{\alpha}\cancel{D}\alpha\lambda^i-\frac{1}{2}(2\cancel{D}\lambda^i+\slashed{Y}^{i\BB}\lambda_\BB )\nonumber\\
&-\frac{i}{4\alpha}\gamma\cdot G\lambda^i+\frac{i}{4}(\gamma\cdot T^{i\BB}\lambda_\BB+\gamma\cdot \check{T}\lambda^i)\bigg)-\frac{8\sigma}{15}\check{D}^{i\AAA}+6 H\cdot T^{i\AAA}-\frac{3i}{2}(\bar{\lambda}^k\gamma\cdot T^{i\AAA}\psi_k\nonumber\\
&+\bar{\lambda}^\BB\gamma\cdot T^{i\AAA}\psi_\BB )=0  \;,\label{5dsigmaiA}  
    \end{align} 
\end{subequations}
}
where 
\begin{align}\label{Homega}
\mathcal{H}_{ab}&=\alpha H_{ab}-\frac{i\alpha}{4}\left(\bar{\lambda}^i\gamma_{ab}\psi_i+\bar{\lambda}^\AAA\gamma_{ab}\psi_\AAA\right)-8\alpha\sigma\check{T}_{ab}\nonumber\\
\omega_{abcd}&=-6G_{[ab}H_{cd]}-\frac{3i}{2}\alpha \left(\bar{\psi}^i\gamma_{[ab}R(Q)_{cd],i}+\bar{\psi}^\AAA\gamma_{[ab}R(Q)_{cd],\AAA}\right)\nonumber\\
(^*\omega)_a&=\frac{1}{4!}\epsilon_{abcde}\omega^{bcde}\;,
\end{align}
where $H_{ab}$ and $G_{ab}$ are the supercovariant field strengths corresponding to the vector gauge fields $A_{\mu}$ and $C_{\mu}$ in the five-dimensional $N=2$ vector multiplet and the dilaton Weyl multiplet respectively. 

The field equation for $\psi^I$ decomposes into the field equations for $\psi^i$ and $\psi^\mathcal{A}$. They are given in \ref{5dpsii} and \ref{5dpsiA}. It can be seen that the $\chi^{ij;\mathcal{A}} $ and $\chi^{i;\mathcal{AB}}$ drop out from the fields equations as a result of the gauge condition \ref{5dggcond} and the fields $\check{\chi}^i$, $\check{\chi}^\mathcal{A}$ can now be solved algebraically using these field equations.  The fields $\chi^{ij;\mathcal{A}} $ and $\chi^{i;\mathcal{AB}}$ remain as independent fields.

Similarly, field equations for $\sigma_{IJ}$ decomposes into the field equations for $\sigma$ and $\sigma^{i\mathcal{A}}$. They are given in \ref{5dsigma} and \ref{5dsigmaiA}. It can be seen that $D^{ij;\mathcal{AB}}$ drops out from the field equations as a result of the gauge condition \ref{5dggcond} and the fields $\check{D}$, $\check{D}^{i\mathcal{A}}$ can now be algebraically solved using these field equations. The field $D^{ij;\mathcal{AB}}$ remains as an independent field. 

 The Maxwell's equation ({\ref{5dMax}}) is re-written by introducing a 3- form $F_{abc}$ dual to $\mathcal{H}_{ab}$ as:
 \begin{align}\label{Fdef}
     F_{abc}=\frac{1}{2}\epsilon_{abcde}\mathcal{H}^{de}\;,
 \end{align}
 where $\varepsilon_{abcde}$ is the standard Levi-Civita defined as $\epsilon_{01234}=1$. In terms of the 3-form $F_{abc}$, the Maxwell's equation becomes
 \begin{align}\label{2formBianchi}
     D_{[a}F_{bcd]}=\frac{1}{4}\omega_{abcd}
 \end{align}
 Now, we solve the above Maxwell's equation by interpreting $F_{abc}$ as the field strength of a 2-form gauge field $B_{\mu\nu}$ and thereby converting the Maxwell's equation into the Bianchi identity of the 2-form gauge field. We then use equations-\ref{Homega} and \ref{Fdef} to solve for $\check{T}_{ab}$ in terms of the field strengths $H_{ab}$ and $F_{abc}$ as shown below    
\begin{align}\label{5dTab}
\check{T}_{ab}&=\frac{1}{4\sigma}H_{ab}-\frac{i}{16\sigma}\bigg(\bar{\lambda}^i\gamma_{ab}\psi_i+\bar{\lambda}^{\mathcal{A}}\gamma_{ab}\psi_{\mathcal{A}}\bigg)+\frac{1}{24\alpha\sigma}\epsilon_{abcde}F^{cde}
\end{align}
The bosonic gauge transformation of the 2-form gauge field $B_{\mu\nu}$ is given as
\begin{align}
  \delta B_{\mu\nu}&=2\partial_{[\mu}\Lambda_{\nu]}+\frac{1}{2}\Lambda^C H_{\mu\nu}+\frac{1}{2}\Lambda^A G_{\mu\nu}
\end{align}
where $\Lambda_\nu$ is the gauge transformation parameter associated with the two-form gauge field $B_{\mu\nu}$ and $\Lambda^C \;, \Lambda^A$ are the gauge transformation parameters associated with the $U(1)$ gauge fields $C_\mu$ and ${A}_{\mu}$ respectively. 
The supersymmetry transformation of this 2-form gauge field is obtained in the same way as we did it in four dimensions. From eq-\ref{Fdef}, we obtain the transformation of the 3-form field strength $F_{abc}$. This, combined with the Bianchi identity (\ref{2formBianchi}), will determine the supersymmetry transformation of the 2-form $B_{\mu\nu}$ which is given in eq-\ref{Btransf5d}. 

We will now summarize the results. The field equations \ref{5dfeqn} render certain auxiliary fields of the original dilaton Weyl multiplet composite in nature. The remaining auxiliary fields, along with the fields from the vector multiplet and some dual gauge fields, merge with the independent gauge fields from the original dilaton Weyl multiplet. This combination forms the new dilaton Weyl multiplet for $N=2$ conformal supergravity in five dimensions, which possesses $SU(2) \times SU(2)$ as its R-symmetry. In table-\ref{newdilaton5d}, we give the full details of this new dilaton Weyl multiplet. 
\begin{table}[t!]
		\centering
		\centering
		\begin{tabular}{|c|c|c|c|}
			\hline
			Field&Properties&$SU(2)\times SU(2)$ Irreps& $w$\\
			\hline
			\multicolumn{4}{|c|}{Independent Gauge fields}\\
			\hline
			$e_\mu^a$&vielbein&\bf{(1,1)}&$-1$\\
			$\psi_\mu^i  $& Gravitino&\bf{(2,1)} &$-1/2$\\
            $ \psi_\mu^{\mathcal{A}} $& Gravitino& \bf{(1,2)}&$-1/2$\\
			$V_\mu^{ij}$&$V_\mu^{ij}=V_\mu^{ji}$, , $SU(2)$ gauge field&\bf{(3,1)}&0\\
            $V_\mu^{\mathcal{A}\mathcal{B}} $&$V_\mu^{\AAA\BB}=V_\mu^{\BB\AAA}$, $SU(2)$ gauge field&\bf{(1,3)}&0\\
			$C_\mu$& $U(1)$ gauge field&\bf{(1,1)}&0\\
            $A_\mu$& $U(1)$ gauge field&\bf{(1,1)}&0\\
            $B_{\mu\nu}$& Tensor gauge field&\bf{(1,1)}&0\\
			\hline
			\multicolumn{4}{|c|}{Covariant fields}\\
			\hline
   $Y_a^{i\mathcal{A}}$&Boson&\bf{(2,2)}&1\\
			$\alpha$& Real Scalar&\bf{(1,1)}&1\\
					$\sigma$&Boson&\bf{(1,1)}&1\\
			$E^{ij}$ & $E^{ij}=E^{ji}$ & \bf{(3,1)} & 1\\
   			$E^{\mathcal{A}\mathcal{B}}$ & $E^{\mathcal{A}\mathcal{B}}=E^{\mathcal{B}\mathcal{A}}$ & \bf{(1,3)} & 1\\
      $E^{i\mathcal{A}}$ & Boson, Lorentz scalar & \bf{(2,2)} & 1\\

			$\lambda^i$&Fermion&\bf{(2,1)}&$1/2$\\
   	$\lambda^{\mathcal{A}} $&Fermion&\bf{(1,2)}&$1/2$\\
       $\psi^i$&Fermion&\bf{(2,1)}&$3/2$\\
   $\psi^{\mathcal{A}}$&Fermion&\bf{(1,2)}&$3/2$\\

			$T^{i\mathcal{A}}_{ab}$  & Boson &\bf{(2,2)}&1\\
			$D^{ij;\mathcal{A}\mathcal{B}}$& $D^{ij;\mathcal{A}\mathcal{B}}=D^{ji;\mathcal{A}\mathcal{B}}=D^{ij;\mathcal{B}\mathcal{A}}$&\bf{(3,3)}&2\\
			$\chi^{ij;\mathcal{A}}$ &$\chi^{ij;\mathcal{A}}=\chi^{ji;\mathcal{A}}$&\bf{(3,2)}&$3/2$\\
   $\chi^{\mathcal{A}\mathcal{B};i}$ &$\chi^{\mathcal{A}\mathcal{B};i}=\chi^{\mathcal{B}\mathcal{A};i}$&\bf{(2,3)}&$3/2$\\
			\hline
		\end{tabular}
		\caption{Five Dimensional New Dilaton Weyl Multiplet}
		\label{newdilaton5d}	
	\end{table}

In \ref{5dsusy}, we give the Q and S supersymmetry transformations of the independent gauge fields of this multiplet:
 \allowdisplaybreaks
{ \begin{subequations}\label{5dsusy}
\begin{align}
\delta e_{\mu}{}^a &= -\frac{1}{2}\left(\bar{\epsilon}^i\gamma^a \psi_{\mu i}+\bar{\epsilon}^\mathcal{A}\gamma^a\psi_{\mu \mathcal{A}} \right)\\[5pt]
\delta\psi_\mu^i&=\mathcal{D}_\mu \epsilon^i-\frac{i}{4}\left(\bar{\psi}^j_{\mu}\gamma^a\lambda_j+\bar{\psi}_\mu^{\mathcal{A}}\gamma^a\lambda_{\mathcal{A}} \right)\gamma_a\epsilon^i+\frac{i}{4}\left(\bar{\epsilon}^j\gamma^a\lambda_j+\bar{\epsilon}^{\mathcal{A}}\gamma^a\lambda_{\mathcal{A}}\right)\gamma_a\psi_\mu^i \nonumber\\
&+\frac{i}{4\alpha}G_\mu{}^a\gamma_a\epsilon^i+\frac{1}{2}\left(\bar{\epsilon}^j\gamma_\mu\lambda_j+\bar{\epsilon}^{\mathcal{A}}\gamma_{\mu}\lambda_{\mathcal{A}}\right)\lambda^i+\frac{i}{2}\left(\bar{\epsilon}^j\psi_{\mu j}+\bar{\epsilon}^\AAA\psi_{\mu \AAA}\right)\lambda^i\nonumber\\
&-\frac{i}{4}\gamma\cdot(\check{T}\gamma_\mu\epsilon^i+ T^{i\mathcal{A}}\gamma_{\mu}\epsilon_{\mathcal{A}})+\frac{1}{2}Y_\mu^{i\mathcal{A}}\epsilon_{\mathcal{A}}+\frac{1}{2}v(\epsilon)^i{}_\mathcal{A}\psi^\mathcal{A}_\mu-\frac{1}{2}v(\psi_\mu)^i{}_\mathcal{A}\epsilon^\mathcal{A}+i\gamma_\mu\eta^i \\[5pt]
\delta\psi_\mu^\AAA&=\mathcal{D}_\mu \epsilon^\AAA-\frac{i}{4}\left(\bar{\psi}^j_{\mu}\gamma^a\lambda_j+\bar{\psi}_\mu^{\mathcal{B}}\gamma^a\lambda_{\mathcal{B}} \right)\gamma_a\epsilon^\AAA+\frac{i}{4}(\bar{\epsilon}^j\gamma^a\lambda_j+\bar{\epsilon}^{\mathcal{B}}\gamma^a\lambda_{\mathcal{B}})\gamma_a\psi_\mu^\AAA \nonumber\\
&+\frac{i}{4\alpha}G_\mu{}^a\gamma_a\epsilon^\AAA+\frac{1}{2}(\bar{\epsilon}^j\gamma_\mu\lambda_j+\bar{\epsilon}^{\mathcal{B}}\gamma_{\mu}\lambda_{\mathcal{B}})\lambda^\AAA+\frac{i}{2}(\bar{\epsilon}^j\psi_{\mu j}+\bar{\epsilon}^\BB\psi_{\mu \BB})\lambda^\AAA\nonumber\\
&+\frac{i}{4}\gamma\cdot(\check{T}\gamma_\mu\epsilon^\AAA+ T^{j\mathcal{A}}\gamma_{\mu}\epsilon_{j})+\frac{1}{2}Y_\mu^{j\mathcal{A}}\epsilon_{j}+\frac{1}{2}v(\epsilon)^\AAA{}_j\psi^j_\mu-\frac{1}{2}v(\psi_\mu)^\AAA{}_j\epsilon^j+i\gamma_\mu\eta^\AAA \\[5pt]
\delta V^{ij}_\mu&=4i\Bar{\epsilon}^{(i}\phi_\mu^{j)} 
+\frac{1}{4}\left(\bar{\epsilon}_{\mathcal{A}} \gamma_\mu \chi^{ij;\mathcal{A}} +\bar{\epsilon}^{(i} \gamma_\mu \check{\chi}^{j)} \right)
+\frac{1}{2} E^{ij}\left(i\bar{\epsilon}^k\psi_{\mu k}+i\bar{\epsilon}^{\mathcal{A}}\psi_{\mu\mathcal{A}} + \bar{\epsilon}^k\gamma_\mu \lambda_k \right.
\nonumber\\
&\left.+\bar{\epsilon}^{\mathcal{A}} \gamma_\mu \lambda_{\mathcal{A}}\right)+v(\epsilon)^{(i}{}_\mathcal{A}\left(Y_\mu^{j)\mathcal{A}}+v(\psi_\mu)^{j)\mathcal{A}} \right) -4i\bar{\eta}^{(i}\psi_\mu^{j)}\\[5pt]
   \delta V^{\mathcal{A}\mathcal{B}}_\mu&=4i\Bar{\epsilon}^{(\mathcal{A}}\phi_\mu^{\mathcal{B})} 
			+\frac{1}{4}\left( \bar{\epsilon}_{i} \gamma_\mu \chi^{\mathcal{A}\mathcal{B};i} -\bar{\epsilon}^{(\mathcal{A}} \gamma_\mu \check{\chi}^{\mathcal{B})} \right)
			+\frac{1}{2} E^{\mathcal{A}\mathcal{B}}\left(i\bar{\epsilon}^k\psi_{\mu k} +i\bar{\epsilon}^{\mathcal{C}}\psi_{\mu\mathcal{C}} + \bar{\epsilon}^k\gamma_\mu \lambda_k
			\right.\nonumber\\
   &\left. + \bar{\epsilon}^{\mathcal{C}} \gamma_\mu \lambda_{\mathcal{C}}\right)+v(\epsilon)^{(\mathcal{A}}{}_i\left(Y_\mu^{i \mathcal{B})}+v(\psi_\mu)^{i\mathcal{B})} \right) -4i\bar{\eta}^{(\mathcal{A}}\psi_\mu^{\mathcal{B})}\\[5pt]
   \delta C_\mu &= -\frac{i}{2}\alpha \left(\bar{\epsilon}^\AAA\psi_{\mu \AAA}+\bar{\epsilon}^i\psi_{\mu i}   \right) 
			-\frac{\alpha}{2}\left( \bar{\epsilon}^i\gamma_\mu \lambda_i+\bar{\epsilon}^\AAA\gamma_\mu \lambda_\AAA \right ) \\[5pt]
   \delta A_{\mu}&=\frac{1}{4}\left(\bar{\epsilon}^i\gamma_\mu\psi_i+\bar{\epsilon}^{\mathcal{A}}\gamma_\mu\psi_{\mathcal{A}}\right) + \frac{1}{2}\sigma\left(\bar{\epsilon^i}\gamma_\mu\lambda_i-\bar{\epsilon}^{\mathcal{A}}\gamma_\mu\lambda_{\mathcal{A}}\right)-\frac{i}{2}\sigma\left( \bar{\epsilon}^i\psi_{\mu,i}-\bar{\epsilon}^{\mathcal{A}}\psi_{\mu,\mathcal{A}} \right)\\[5pt]
\delta B_{\mu\nu}&=\frac{i}{4}\alpha\left( \bar{\epsilon}^i\gamma_{\mu\nu}\psi_i+ \bar{\epsilon}^\mathcal{A}\gamma_{\mu\nu}\psi_\mathcal{A}\right)+\alpha\sigma\left( \bar{\epsilon}^i\gamma_{[\mu}\psi_{\nu]i}-\bar{\epsilon}^\mathcal{A}\gamma_{[\mu}\psi_{\nu]\mathcal{A}}\right)-A_{[\mu}\delta(\epsilon)C_{\nu]}\nonumber\\
&-C_{[\mu}\delta(\epsilon)A_{\nu]} \label{Btransf5d}\end{align}
\end{subequations}
In \ref{5dsusycov}, we give the Q and S supersymmetry transformations of the covariant fields of this multiplet:
\begin{subequations}\label{5dsusycov}

\end{subequations}} 
\section{The $(2,0)$ dilaton Weyl multiplet in six dimensions}\label{6dnewdilaton}
In six dimensions, there is no known dilaton Weyl multiplet for $(2,0)$ conformal supergravity. There is a $(2,0)$ standard Weyl multiplet that is known \cite{Bergshoeff:1999db}. In this section we give the details of the construction of a dilaton Weyl multiplet for $(2,0)$ conformal supergravity in six dimensions. In order to do this, we couple a $(2,0)$ tensor multiplet multiplet to the standard Weyl multiplet given in \cite{Bergshoeff:1999db}. The field equations of the tensor multiplet are then used to solve for some components of the auxiliary fields of the standard Weyl multiplet. Similar to the case of four and five dimensions, in order to solve the tensor multiplet field equations, we first need to break the $USp(4)$ R-symmetry of the standard Weyl multiplet to $SU(2)\times SU(2)$. The decompositions of the fields of the standard Weyl multiplet and the tensor multiplet carrying $USp(4)$ irreducible representations into objects carrying $SU(2)\times SU(2)$ irreducible representations is given in appendix \ref{6dirrepsbreaking}. We use a notation similar to the case of four as well as five dimensions.

The field $\phi^{IJ}$ of the tensor multiplet which is in the $\bf{5}$ of $USp(4)$ decomposes into $\phi^{i\mathcal{A}}$ in the $\bf{(2,2)}$ and $\phi $ in the  $\bf{(1,1)}$ of $SU(2)\times SU(2)$. We gauge fix $USp(4)$ to $SU(2)\times SU(2)$ by using the gauge fixing condition:
\begin{align}\label{6dggcond}
\phi^{i\mathcal{A}}=0     
\end{align}

 Q-supersymmetry will not preserve this gauge fixing condition, and hence the unbroken Q-supersymmetry will have to be appropriately redefined as shown below:
\begin{align}\label{compensator}
\delta_Q(\epsilon)=\delta_Q^{\text{old}}(\epsilon)+\delta_{USp(4)}(\Lambda^{i\mathcal{A}}=w(\epsilon)^{i\AAA})
	\end{align}
	
where $w(\epsilon)^{i\AAA}$ is an $\epsilon$ and field dependent $USp(4)$ parameter which is obtained by demanding that $\delta_Q\phi^{i\mathcal{A}}=0$ and is given as:
\begin{align}
w(\epsilon)^{i\mathcal{A}}=2\phi^{-1}\bigg(\bar{\epsilon}^{i}\psi^{\mathcal{A}}-\bar{\epsilon}^\mathcal{A}\psi^i\bigg)
\end{align}

The $USp(4)$ gauge field decompose into objects carrying  representations as shown in Table-\ref{6dbreak}. Similar, to the case of four dimensions, $V_\mu^{ij} $ and $V_\mu^{\AAA\BB} $ become the $SU(2) \times SU(2)$ gauge fields, whereas the $V_\mu^{i\AAA}$ will have to be modified by adding explicit gravitino terms to convert it into a covariant matter field $Y^{i\AAA}_a$ and the relation is given below:

\begin{align}
    Y_a^{i\AAA}&=V_a^{i\AAA}-w(\psi_a)^{i\AAA}
\end{align}

The supercovariant derivative $D_a$ of a field appearing in the standard Weyl multiplet is supercovariant w.r.t the old supersymmetry as well as $USp(4)$ R-symmetry. For the dilaton Weyl multiplet, it will decompose into a supercovariant derivative $D_a^{\text{new}}$, which is covariant with respect to the new supersymmetry and $SU(2)\times SU(2)$ R-symmetry, along with some supercovariant terms involving $Y_a^{i\mathcal{A}}$. For instance, for $\lambda^i, \lambda^\mathcal{A}$, we have
\begin{align}
D_a \lambda^i&=D^{\text{new}}_a\lambda^i+\frac{1}{2}Y_a^{i\mathcal{A}}\lambda_\mathcal{A} \nonumber\\
D_a\lambda^\mathcal{A}&=D_a^{\text{new}} \lambda^\mathcal{A}+\frac{1}{2}Y_a^{j\mathcal{A}}\lambda_j
\end{align}
Now, we first rewrite the field equations of the tensor multiplet \cite{Bergshoeff:1999db} after decomposing them into $SU(2) \times SU(2)$ representations and using the gauge fixing condition (\ref{6dggcond}):
\begin{subequations}\label{6dreplace}
\begin{align}
&\check{T}_{abc}=\frac{1}{2\phi}H^-_{abc}\;,\label{6dmaxwell} \\[5pt]
&\cancel{D}\psi^i+\frac{1}{2}\slashed{Y}^{i\mathcal{A}}\psi_{\mathcal{A}}-\frac{1}{12}\gamma\cdot \check{T} \psi^i-\frac{1}{2}\gamma \cdot T^{i\mathcal{A}}\psi_\mathcal{A}-\frac{4\phi}{15}\check{\chi}^i=0\;, \label{6dzetai}\\[5pt]
&\cancel{D}\psi^\AAA+\frac{1}{2}\slashed{Y}^{j\mathcal{A}}\psi_j+\frac{1}{12}\gamma\cdot \check{T} \psi^\AAA+\frac{1}{2}\gamma \cdot T^{i\mathcal{A}}\psi_i-\frac{4\phi}{15}\check{\chi}^\AAA=0\;, \label{6dzetaA}\\[5pt]
&D^a D_a\phi-\frac{1}{2}Y_a^{j\BB}Y^a_{j\BB}-\frac{4\phi}{15}\check{D} +\frac{1}{3}H^+_{abc}\check{T}^{abc}+\frac{16}{15}(\check{\bar{\chi}}^i\psi_i+\check{\bar{\chi}}^\AAA\psi_\AAA )=0 \;,\label{6dphieq}\\[5pt]
&-\phi D^aY_{a,i\AAA}-2Y_{a,i\AAA}D^a\phi -\frac{4\phi}{15}\check{D}_{i\AAA}+\frac{1}{3}H^+_{abc}\check{T}^{abc}_{i\AAA}+\frac{8}{15}(-2\bar{\chi}_{ik;\AAA}\psi^k+\check{\bar{\chi}}_\AAA\psi_i+2\bar{\chi}_{\AAA\BB;i}\psi^\BB \nonumber\\
&-\check{\bar{\chi}}_i\psi_{\AAA} ) =0\;, \label{6dphiiA}
\end{align}
\end{subequations}
The field equation for the tensor gauge field $B_{\mu\nu}$ (\ref{6dmaxwell}) is a little peculiar in contrast to the Maxwell's equations in four and five dimensions. This equation is already algebraic, and we do not need to introduce any dual gauge field to solve it. This algebraic equation determines the auxiliary field $\check{T}_{abc}$ in terms of the field strength of the two-form gauge field $B_{\mu\nu}$. The components $T^{i\AAA}_{abc}$ drops out from the field equation due to the gauge condition (\ref{6dggcond}) and remains an independent field.  

The solution for the remaining equations is very similar to the case of four and five dimensions. The field equation for $\zeta^I$ decomposes into the field equations for $\zeta^i$ and $\zeta^\mathcal{A}$. They are given in (\ref{6dzetai}) and (\ref{6dzetaA}). It can be seen that the fields $\chi^{ij;\mathcal{A}} $ and $\chi^{i;\mathcal{AB}}$ drop out from the fields equations as a result of the gauge condition (\ref{6dggcond}) and the fields $\check{\chi}^i$, $\check{\chi}^\mathcal{A}$ can now be solved algebraically using these field equations and are completely determined in terms of the other fields.  The fields $\chi^{ij;\mathcal{A}} $ and $\chi^{i;\mathcal{AB}}$ remain as independent fields.

Similarly, field equations for $\phi_{IJ} $ decomposes into the field equations for $\phi$ and $\phi^{i\mathcal{A}}$. They are given in (\ref{6dphieq}) and (\ref{6dphiiA}). It can be seen that $D^{ij;\mathcal{AB}}$ drops out from the field equations as a result of the gauge condition (\ref{6dggcond}) and the fields $\check{D}$, $\check{D}^{i\mathcal{A}}$ can now be solved algebraically using these field equations. The field $D^{ij;\AAA\BB}$ remains as an independent field.

\begin{table}[t!]
		\centering
		\centering
		\begin{tabular}{ |c|c|c|c| }
			\hline
			Field&Properties&$SU(2)\times SU(2)$ Irreps&$w$\\
			\hline
			\multicolumn{4}{|c|}{Independent Gauge fields}\\
			\hline
			$e_\mu{}^a$&vielbein &\bf{(1,1)}&$-1$\\
			$\psi_\mu^i $& $\Gamma_*\psi_\mu^i=\psi_\mu^i$, gravitino &\bf{(2,1)} &$-1/2$\\
            $ \psi_\mu^{\mathcal{A}}$&  $\Gamma_*\psi_\mu^{\mathcal{A}}=\psi_\mu^{\mathcal{A}}$, gravitino & \bf{(1,2)}&$-1/2$\\
			$V_\mu^{ij} $&$V_\mu^{ij}=V_\mu^{ji}$, $SU(2)$ gauge field&\bf{(3,1),(1,3)}&0\\
            $ V_\mu^{\mathcal{A}\mathcal{B}}$&$ V_\mu^{\mathcal{A}\mathcal{B}}=V_\mu^{\mathcal{B}\mathcal{A}}$, $SU(2)$ gauge field&\bf{(3,1),(1,3)}&0\\
			$B_{\mu \nu}$  &Tensor gauge field&\bf{(1,1)}&0\\
                     \hline
			\multicolumn{4}{|c|}{Covariant fields}\\
			\hline
			$Y_a^{i\mathcal{A}}$&Boson&$\bf{(2,2)}$&1\\
			$T_{abc}^{i\mathcal{A}}$& Totally anti-symmetric \& anti-self-dual&$\bf{(2,2)}$&1\\
			$\phi$&Scalar&$\bf{(1,1)}$&2\\
           
			$D^{ij;\mathcal{A}\mathcal{B}}$&$D^{ij;\mathcal{A}\mathcal{B}}=D^{ji;\mathcal{A}\mathcal{B}}=D^{ij;\mathcal{B}\mathcal{A}}$ &$\bf{(3,3)}$&1\\
			$\psi^i$ & $\gamma_* \psi^i=-\psi^i$&$\bf{(2,1)}$&$5/2$\\
			$\psi^{\mathcal{A}}$ & $\gamma_* \psi^{\mathcal{A}}=-\psi^{\mathcal{A}}$&$\bf{(1,2)}$&$5/2$\\
			$\chi^{ij;\mathcal{A}}$&$\chi^{ij;\mathcal{A}}=\chi^{ji;\mathcal{A}}$, $\Gamma_*\chi^{ij;\mathcal{A}}=\chi^{ij;\mathcal{A}}$&$\bf{(3,2)}$&$3/2$\\
				$\chi^{\mathcal{A}\mathcal{B};i}$&$\chi^{\mathcal{A}\mathcal{B};i}=\chi^{\mathcal{B}\mathcal{A};i}$, $\Gamma_*\chi^{\mathcal{A}\mathcal{B};i}=\chi^{\mathcal{A}\mathcal{B};i}$&$\bf{(2,3)}$&$3/2$\\
    
			\hline
		\end{tabular}
		\caption{Six-Dimensional Dilaton Weyl Multiplet}
		\label{6ddil}	
	\end{table}
Now, we summarise the results. The field equations \ref{6dreplace} make some of the auxiliary fields of the standard Weyl multiplet composite. The remaining auxiliary fields, together with the fields of the tensor multiplet and all the remaining gauge fields of the standard Weyl multiplet, combine into a new off-shell multiplet, which is a $(2,0)$ dilaton Weyl multiplet where the R-symmetry is $SU(2)\times SU(2)$. In Table-\ref{6ddil}, we give the full details of this $(2,0)$ dilaton Weyl multiplet.

The Q and S supersymmetry transformations of the independent gauge fields are given in equations-\ref{6dtransf}.
{\allowdisplaybreaks
\begin{subequations}\label{6dtransf}
\begin{align}
\delta e_\mu{}^a&=\frac{1}{2}(\bar{\epsilon}^i\gamma^a\psi_{\mu i}+\bar{\epsilon}^\AAA\gamma^a\psi_{\mu \AAA}) \\[5pt]
\delta\psi_\mu^i&=\mathcal{D}_\mu\epsilon^i+\frac{1}{2}Y_\mu^{i\AAA}\epsilon_\AAA+\frac{1}{2}u(\psi_\mu)^{i\AAA}\epsilon_\AAA +\frac{1}{24}\gamma \cdot (\check{T}\gamma_\mu\epsilon^i+T^{i\AAA}\gamma_\mu\epsilon_\AAA)-\frac{1}{2}w(\epsilon)^{i\AAA}\psi_{\mu \AAA}\nonumber\\
&+\gamma_\mu\eta^i \\[5pt]
\delta\psi_\mu^{\mathcal{A}}&=\mathcal{D}_\mu\epsilon^{\mathcal{A}}+\frac{1}{2}Y_\mu^{j\AAA}\epsilon_j+\frac{1}{2}u(\psi_\mu)^{j\AAA}\epsilon_j -\frac{1}{24}\gamma \cdot (\check{T}\gamma_\mu\epsilon^\AAA +T^{j\AAA}\gamma_\mu\epsilon_j)-\frac{1}{2}w(\epsilon)^{j\AAA}\psi_{\mu j}\nonumber\\
&+\gamma_\mu\eta^\AAA \\[5pt]
 	{\delta} V_\mu^{ij}&=-4\bar{\epsilon}^{(i}\phi^{j)}_\mu-\frac{4}{15}(\bar{\epsilon}_\AAA\gamma_\mu\chi^{(ij);\AAA}+\bar{\epsilon}^{(i}\gamma_\mu\check{\chi}^{j)})+w(\epsilon)^{(i}{}_\AAA Y^{j)\AAA}_\mu+w(\epsilon)^{(i}{}_\AAA w(\psi_\mu)^{j)\AAA}\nonumber\\
    &-4\bar{\eta}^{(i}\psi_\mu^{j)}\\[5pt]
 	\delta V_{\mu}^{\mathcal{A}\mathcal{B}}&=-4\bar{\epsilon}^{(\AAA}\phi^{\BB)}_\mu+\frac{4}{15}(-\bar{\epsilon}_k\gamma_\mu\chi^{k;\AAA\BB}+\bar{\epsilon}^{(\AAA}\gamma_\mu\check{\chi}^{\BB)})-w(\epsilon)^{j(\AAA} Y_{\mu j}{}^{\BB)}-w(\epsilon)^{j(\AAA} w(\psi_\mu)_j{}^{\BB)}\nonumber\\
    &-4\bar{\eta}^{(\AAA}\psi_\mu^{\BB)}\\[5pt]
 	{\delta}B_{\mu \nu}&= -\bar{\epsilon}^i\gamma_{\mu\nu}\psi_i-\bar{\epsilon}^{\mathcal{A}}\gamma_{\mu\nu}\psi_{\mathcal{A}}-(\bar{\epsilon}^i\gamma_{[\mu}\psi_{\nu]i}-\bar{\epsilon}^{\mathcal{A}}\gamma_{[\mu}\psi_{\nu]\mathcal{A}})\phi
    \end{align}
        \end{subequations}
The Q and S supersymmetry transformations of the covariant fields are given in equations-\ref{6dtransfcov}.        
        \begin{subequations}\label{6dtransfcov}
\begin{align}
 	\delta Y_a^{i \mathcal{A}}&=D_aw(\epsilon)^{i\AAA}+ \frac{4}{15}(\bar{\epsilon}_j\gamma_a\chi^{ij;\AAA}+\bar{\epsilon}_\BB\gamma_a \chi^{\AAA\BB;i}-2\bar{\epsilon}^i\gamma_a \check{\chi}^\AAA+2\bar{\epsilon}^\AAA\gamma_a \check{\chi}^i )\nonumber\\
    &+\frac{1}{12\phi}\{\bar{\psi}^\AAA\gamma\cdot (\check{T}\gamma_a \epsilon^i+T^{i\BB}\gamma_a\epsilon_\BB )+\bar{\psi}^i\gamma\cdot(\check{T}\gamma_a \epsilon^\AAA+T^{j\AAA}\gamma_a \epsilon_j )\}+w(\gamma_a\eta)^{i\AAA}   \\[5pt]
{\delta}\psi^i&=\frac{1}{48}H_{abc}^{+}\gamma^{abc}\epsilon^i+\frac{1}{4}\cancel{D}\phi \epsilon^i+\frac{1}{2}w(\epsilon)^i{}_\AAA\psi^\AAA -\phi\eta^i\\[5pt]
{\delta}\psi^\AAA&=\frac{1}{48}H_{abc}^{+}\gamma^{abc}\epsilon^\AAA-\frac{1}{4}\cancel{D}\phi \epsilon^\AAA+\frac{1}{2}w(\epsilon)^\AAA{}_j\psi^j +\phi\eta^\AAA\\[5pt]
\delta \phi&=\bar{\epsilon}^i\psi_i+\bar{\epsilon}^{\mathcal{A}}\psi_{\mathcal{A}}\\[5pt]
\delta T^{i\mathcal{A}}_{abc}&=\frac{1}{16}\bar{\epsilon}^i\gamma^{de}\gamma_{abc}R(Q)_{de}^{\mathcal{A}}+\frac{1}{15}\bar{\epsilon}_j\gamma_{abc}\chi^{ij;\mathcal{A}}+\frac{1}{30}\bar{\epsilon}^i\gamma_{abc}\check{\chi}^{\mathcal{A}}-\frac{1}{16}\bar{\epsilon}^{\mathcal{A}}\gamma^{de}\gamma_{abc}R(Q)_{de}^i\nonumber\\
&-\frac{1}{15}\bar{\epsilon}_{\mathcal{B}}\gamma_{abc}\chi^{i;\mathcal{A}\mathcal{B}}+\frac{1}{30}\bar{\epsilon}^{\mathcal{A}}\gamma_{abc}\check{\chi}^i+w(\epsilon)^{i\AAA}\check{T}_{abc}  \\[5pt]
\delta\chi^{ij;\mathcal{A}}&=\frac{5}{32}\gamma^{abc}\gamma^d\epsilon^{(i}D_dT^{j)\mathcal{A}}_{abc}-\frac{5}{32}\gamma\cdot\check{T}\gamma^d\epsilon^{(i}Y_d^{j)\AAA}  -\frac{15}{32}\left(\gamma\cdot R(V)^{ij} -\gamma^{ab}Y_a^{(i}{}_{\mathcal{B}}Y_b^{j)\mathcal{B}}\right)\epsilon^{\mathcal{A}}
\nonumber\\
&+\frac{15}{16}\gamma^{ab}\epsilon^{(i} \left(2D_aY^{j)\mathcal{A}}_b +w(R(Q)_{ab})^{j)\AAA}\right)+\frac{1}{4}\epsilon^{(i}\check{D}^{j)\AAA} +\frac{1}{4}{D}^{ij;\mathcal{A}\mathcal{B}}\epsilon_{\mathcal{B}}+\frac{5}{4}\check{\chi}^{(i}w(\epsilon)^{j)\AAA} \nonumber\\
&-\frac{1}{2}w(\epsilon)^{(j}{}_\BB \chi^{i);\AAA\BB}+\frac{5}{8}\gamma^{abc}\eta^{(i}T^{j)\mathcal{A}}_{abc}\\[5pt]
\delta\chi^{i;\mathcal{A}\mathcal{B}}&=-\frac{5}{32}D^dT^{abc i(\mathcal{A}}\gamma_{abc}\gamma_d\epsilon^{\mathcal{B})}+\frac{5}{32}Y^{d i(\mathcal{A}}\gamma\cdot \check{T}\gamma_d\epsilon^{\mathcal{B})} -\frac{15}{32}\gamma\cdot R(V)^{\mathcal{A}\mathcal{B}}\epsilon^{i}\nonumber\\
&-\frac{15}{32}\gamma^{ab}Y_a^{j(\mathcal{A}}Y_b^{\mathcal{B})}{}_j\epsilon^i+\frac{15}{16}\gamma^{ab}\left(2D_aY^{i(\mathcal{A}}_b +w(R(Q)_{ab})^{i(\AAA} \right)\epsilon^{\mathcal{B})}+\frac{1}{4}\check{D}^{i(\AAA}\epsilon^{\BB)} \nonumber\\
&+\frac{1}{4}D^{ij;\mathcal{A}\mathcal{B}}\epsilon_{j}-\frac{5}{4}w(\epsilon)^{i(\AAA} \check{\chi}^{\mathcal{B})}-\frac{1}{2}\chi^{ij;(\mathcal{A}} w(\epsilon)_j{}^{\BB)}-\frac{5}{8}\gamma\cdot T^{i(\mathcal{A}}\eta^{\mathcal{B})}\\[5pt]
\delta D^{ij;\mathcal{A}\mathcal{B}}&=\frac{1}{2}(\bar{\epsilon}^i\cancel{D}\chi^{\mathcal{A}\mathcal{B};j}+\bar{\epsilon}^{\mathcal{A}}\cancel{D}\chi^{ij;\mathcal{B}})+\frac{5}{8}\left(\bar{\epsilon}^i\slashed{Y}^{j\mathcal{A}}\chi^{\mathcal{B}}-\bar{\epsilon}^{\mathcal{A}}\slashed{Y}^{j\mathcal{B}}\chi^{i}\right)\nonumber\\
&+\frac{1}{4}\left(\bar{\epsilon}^i\slashed{Y}_{k}{}^{\mathcal{B}}\chi^{jk;\mathcal{A}}+\bar{\epsilon}^{\mathcal{A}}\slashed{Y}^{j}{}_{\mathcal{C}}\chi^{\mathcal{B}\mathcal{C};i}\right)+w(\epsilon)^{i\AAA}\check{D}^{j\BB} -\bar{\eta}^i\chi^{\mathcal{A}\mathcal{B};j}-\bar{\eta}^{\mathcal{A}}\chi^{ij;\mathcal{B}}\nonumber\\
&+\left(i\leftrightarrow j+\mathcal{A}\leftrightarrow \mathcal{B}\right)
	\end{align}
        \end{subequations}}
 \section{Discussion}\label{conclusion}
 Weyl multiplets are the gauge multiplets of conformal supergravity. They carry the gauge fields of conformal supergravity together with some covariant matter fields. In various cases, the Weyl multiplets have been known to exist in two variants, i.e. standard and dilaton, which have the same gauge field contents but different covariant field contents. The dilaton Weyl multiplet for $(2,0)$ conformal supergravity in six dimensions was lacking, which we constructed in this paper. We also showed that in the case of $N=4$ and $N=2$ conformal supergravity in four and five dimensions respectively, we can even have a different variant of the dilaton Weyl multiplet. These multiplets contain different covariant field contents compared to the previously constructed dilaton Weyl multiplet. They all have one thing in common i.e. the R-symmetry realized on them is $SU(2)\times SU(2)$. We expect that the dilaton Weyl multiplets constructed in this paper in four, five, and six dimensions will be related to each other by dimensional reductions on a circle or a 2-torus. This is currently a work in progress. 

It would be interesting to see if one can keep adding compensating multiplets to the dilaton Weyl multiplets and keep generating newer dilaton Weyl multiplets until one saturates all the compensating multiplets. If one can obtain a dilaton Weyl multiplet with the maximum number of compensating multiplets, it would probably mean that the dilaton Weyl multiplet would be enough for the construction of Poincar{\'e} supergravity without the requirement of any more compensating multiplet, as the dilaton Weyl multiplet will have in-built fields to compensate for the extra symmetries in order to go from conformal supergravity to Poincar{\'e} supergravity. In that case, the Poincar{\'e} supergravity would be completely off-shell as one would not be using any on-shell compensating multiplets. 

As seen earlier in \cite{Ciceri:2024xxf} as well as in this paper, in order to get dilaton Weyl multiplets for maximal conformal supergravity theories, one needs to break the R-symmetry. In \cite{Ciceri:2024xxf}, the $SU(4)$ R-symmetry of the standard Weyl multiplet was broken to $USp(4)$ and in the current work, the $USp(4)$ R-symmetry was further broken to $SU(2)\times SU(2)$. In one of the upcoming works, which involves the authors of this paper, this breaking of R-symmetry is also required in the construction of dilaton Weyl multiplets in a non-maximal conformal supergravity theory such as $N=3$ conformal supergravity in four dimensions. However, such a breaking of R-symmetry was not necessary for the construction of dilaton Weyl multiplets for $N=2$ conformal supergravity in four dimensions or $N=1$ conformal supergravity in five dimensions or $(1,0)$ conformal supergravity in six dimensions. The reason behind whether there is a requirement to break the R-symmetry or not has to primarily do with the structure of the auxiliary $D$ and $\chi$ fields in the Weyl multiplet. If the representation carried by these fields is such that they decouple from the rest of the fields in the equations of motion of the compensating multiplet, then breaking of R-symmetry is not required, which is the case for $N=2$ conformal supergravity in four dimensions or $N=1$ conformal supergravity in five dimensions or $(1,0)$ conformal supergravity in six dimensions. However, if they don't decouple, then we need to break the R-symmetry so that some components of the auxiliary $D$ field and the $\chi$ field drop out of the equations of motion of the compensating multiplet and the remaining components decouple in the equations of motion. This is what we saw in this paper and \cite{Ciceri:2024xxf} as well is in one of our upcoming papers on the construction of dilaton Weyl multiplets in $N=3$ conformal supergravity. This observation will be crucial in determining how far we can push this program of coupling more and more compensating multiplets and generating more and more dilaton Weyl multiplets. We hope to address these issues in the future.

It seems that dilaton Weyl multiplets are a gateway to construct conformal supergravity theory in cases where there are no rigid conformal superalgebras as per Nahm's classification. The example of $N=2$ conformal supergravity theory in five dimensions is one such case. Another example is that of $N=1$ conformal supergravity in ten dimensions \cite{Bergshoeff:1982az}. Recently, the paper \cite{Hahner:2024hak} has discussed the existence of conformal supergravity theory, in an example independent way in any dimensions and with any amount of supersymmetry, even when a rigid superconformal algebra does not exist and seems to reproduce all the known examples. Thus, it seems plausible that one can construct conformal supergravity theory in any dimension and with any amount of supersymmetry. It would be interesting to see if the approach of superconformal multiplet calculus can be used to construct Poincar{\'e} supergravity theories in all such cases. Another interesting example where one might be able to construct a conformal supergravity theory, in the absence of a rigid superconformal algebra, using a dilaton Weyl multiplet would be a $(1,1)$ theory in six dimensions. This theory will have an $SU(2)\times SU(2)$ R-symmetry and would give rise to five-dimensional conformal supergravity with $SU(2)\times SU(2)$ R-symmetry upon dimensional reduction on a circle. It would be interesting to see whether it gives us the same five-dimensional theory constructed in this paper or a different one. This is currently a work in progress. 

As discussed earlier, conformal supergravity coupled to a certain number of compensating multiplets is gauge equivalent to Poincar{\'e} supergravity. The number of compensating multiplets will depend on what Weyl multiplet we are using. For example, in the case of $N=4$ supergravity in four dimensions, if we start with a standard Weyl multiple, one needs 6 compensating vector multiplets \cite{deRoo:1984zyh} and as speculated in \cite{Ciceri:2024xxf}, if one starts with the old dilaton Weyl multiplet one would need 5 compensating vector multiplets. The Poincar{\'e} supergravity constructed from both the Weyl multiplets will be very different in the way the duality symmetry $SU(1,1)$ acts on the fields as $SU(1,1)$ is realized off-shell on the components of the Weyl multiplet whereas it is realized on-shell, i.e., on the equations of motion of the compensating vector multiplets. However, perhaps the Poincar{\'e} supergravity obtained from both the Weyl multiplets is not inequivalent. Perhaps the Poincar{\'e} supergravity constructed from the old dilaton Weyl multiplet coupled to $5$ compensating vector multiplet is equivalent to a matter-coupled Poincar{\'e} supergravity constructed from the standard Weyl multiplet coupled to $6+1$ vector multiplet (6 compensating and 1 physical vector multiplet) in a different symplectic frame.\footnote{We thank Ashoke Sen for some illuminating discussions in this context.} The above speculation is primarily based on counting the degrees of freedom of the Poincar{\'e} supergravity in both cases. One can count the number of physical scalars and vectors in the Poincar{\'e} supergravity theory, which turns out to be $6$ and $7$, respectively, in both cases. A detailed construction of Poincar{\'e} supergravity using both the Weyl multiplets would shed more light on this. In the same way one would expect that to arrive at a Poincar{\'e} supergravity using the new dilaton Weyl multiplet constructed in this paper for $N=4$ conformal supergravity in four dimensions, one would need to couple it to a minimum of $4$ vector multiplets and just based on counting the number of physical degrees of freedom, the resulting theory might be equivalent to a Poincar{\'e} supergravity obtained from the standard Weyl multiplet coupled to $6+2$ vector multiplets or from the old dilaton Weyl multiplet coupled to $5+1$ vector multiplets in a different symplectic frame.\footnote{The number of physical scalars and vectors in each of the cases turns out to be $12$ and $8$, respectively.} We hope to address these issues in the future.

\acknowledgments
We thank Aravind Aikot, Daniel Butter, Franz Ciceri, Subramanya Hegde, Alok Laddha, Madhu Mishra, Ashoke Sen, and Amitabh Virmani for useful discussions. We would also like to thank the anonymous referee of \cite{Adhikari:2023tzi} for their useful comments, which made us think about this work. SA would like to thank Chennai Mathematical Institute, Harish Chandra Research Institute, Banaras Hindu University, IIT Bombay (and the organizers for ST4 2024), and IISER Pune for their hospitality during the course of this work. BS would like to thank Harish Chandra Research Institute, IISER Pune, Institute of Mathematical Sciences, Chennai Mathematical Institute, and International Center for Theoretical Sciences for their hospitality during the course of the work.
			
			\appendix
 \section{Relations between $USp(4)$ and $SU(2)\otimes SU(2) $ irreps} 
 \subsection{Four dimensions}\label{4dirrepsbreaking}
\begin{table}[h!]
	\centering
	\centering
	\begin{tabular}{ |p{4cm}|p{6cm}|}
		\hline
		$USp(4)$ Irreps & $SU(2)\times SU(2)$\\
			\hline
		\multicolumn{2}{|c|}{Independent fields of the old dilaton Weyl multiplet}\\
		\hline
$e_\mu^a(\bf{1})$&$e_\mu^a(\bf{1,1})$\\
  $\psi_\mu^I(\bf{4})$ & $\psi_\mu^i(\bf{2,1}),\psi_\mu^{\mathcal{A}}$\\
  $V_\mu^{IJ}(\bf{10}) $ & $V_\mu^{ij}(\bf{3,1})$$,V_\mu^{\mathcal{AB}}(\bf{1,3})$$, Y_a^{i\mathcal{A}}(\bf{2,2}) $\\ 
  $\mathcal{A}_{\mu;\alpha} (\bf{1}) $ & $\mathcal{A}_{\mu;\alpha}(\bf{1,1}) $\\
$X_a^{IJ} (\bf{5}) $& $\check{X}_a(\bf{1,1}) $, $X_a^{i\mathcal{A}}(\bf{2,2}) $\\
$\rho (\bf{1})$ & $\rho (\bf{1,1})$\\
$\phi_\alpha $ & $\phi_\alpha(\bf{1,1}) $\\
$E_{IJ}(\bf{10}) $ & $E_{ij}(\bf{3,1}) $, $E_{\mathcal{A}\mathcal{B}}(\bf{1,3})$, $E_{i\mathcal{A}}(\bf{2,2}) $\\
$\Lambda_I(\bf{4}) $& $\Lambda_i(\bf{2,1})$, $\Lambda_\mathcal{A}(\bf{1,2}) $\\
$\psi_I(\bf{4}) $& $\psi_i(\bf{2,1})$, $\psi_\mathcal{A}(\bf{1,2}) $\\
$T_{ab}^{IJ}$  $(\bf{5})$& $\check{T}^{-}_{ab}$ $(\bf{1,1})$,$T_{ab}^{i\mathcal{A}}$  $(\bf{2,2})$\\
		$\chi_I{}^{JK}$  $(\bf{16})$&$\check{\chi}_i$ $(\bf{2,1})$, $\check{\chi}_\mathcal{A}$ $(\bf{1,2})$, $\chi^{ij;\mathcal{A}}$
  $(\bf{3,2})$, $\chi^{i;\mathcal{A}\mathcal{B}}$ $(\bf{2,3})$\\ 
$D^{IJ}{}_{KL}$ $ (\bf{14})$&$\check{D}$ $(\bf{1,1})$, $\check{D}^{i\mathcal{A}}$ $(\bf{2,2})$, $D^{ij;\mathcal{A}\mathcal{B}}$ $(\bf{3,3})$\\
		\hline
		\multicolumn{2}{|c|}{Fields of the vector multiplet}\\
		\hline
$W_{\mu }$ $(\bf{1})$& $W_{\mu }$ $(\bf{1,1})$\\
			$\zeta^I$ $(\bf{4})$& $\zeta^i$ $(\bf{2,1})$, $\zeta^\mathcal{A}$ $(\bf{1,2})$\\
			$\phi_{IJ}$ $(\bf{5})$ & $\phi$ $(\bf{1,1})$, $\phi_{i\mathcal{A}}$ $(\bf{2,2})$\\
   $\sigma(\bf{1})$ & $\sigma(\bf{1,1}) $\\
			\hline
	\end{tabular}
	\caption{Decomposition of $USp(4)$ irreps to $SU(2)\times SU(2)$ in four dimensions}
	\label{4dbreak}	
\end{table}
 In table \ref{4dbreak}, we give the details of how the $USp(4)$ irreducible representations carried by the fields of the old dilaton Weyl multiplet and vector multiplet in $N=4$ conformal supergravity in four dimensions decompose into $SU(2)\times SU(2) $ irreducible representations. The relations between the fields appearing on the L.H.S of the table and the R.H.S of the table are as given below. We denote some of the components of the fields of the old dilaton Weyl multiplet on the R.H.S of the table, which becomes dependent in the new dilaton Weyl multiplet, with an inverted hat notation.
\allowdisplaybreaks
{\begin{align}\label{4d2SU(2)}
&V_{\mu}^{ij}=V_{\mu}^{ij}\;, V_{\mu}^{\mathcal{A}\mathcal{B}}=V_{\mu}^{\mathcal{A}\mathcal{B}}\;, V_{a}^{i\mathcal{A}}-\frac{1}{2}u(\psi_a)^{i\mathcal{A}}=Y_{a}^{i\mathcal{A}}\nonumber \\
&X_a^{ij}=i\varepsilon^{ij} \check{X}_a\;, X_a^{\mathcal{A}\mathcal{B}}=i\check{X}_a \varepsilon^{\mathcal{A}\mathcal{B}}\;, X_{a}^{i\mathcal{A}}=X_{a}^{i\mathcal{A}} \nonumber \\ 
&T_{ab}^{ij}=\varepsilon^{ij}\check{T}^-_{ab}\;, T^{\mathcal{A}\mathcal{B}}_{ab}=\varepsilon^{\mathcal{A}\mathcal{B}}\check{T}_{ab}^-\;, T_{ab}^{i\mathcal{A}}=T_{ab}^{i\mathcal{A}} \nonumber \\
&\chi_i{}^{jk}=\varepsilon^{jk}\check{\chi}_i, \chi_i{}^{\mathcal{A}\mathcal{B}}=\varepsilon^{\mathcal{A}\mathcal{B}}\check{\chi}_i\;, \chi_{\mathcal{A}}{}^{jk}=\varepsilon^{jk}\check{\chi}_\mathcal{A}\;, \chi_{\mathcal{A}}{}^{\mathcal{B}\mathcal{C}}=\varepsilon^{\mathcal{B}\mathcal{C}}\check{\chi}_\mathcal{A}\nonumber\\
&\chi_i{}^{j\mathcal{A}}=\varepsilon_{ik}\chi^{jk;\mathcal{A}} +\frac{1}{2}\delta_i^j \varepsilon^{\mathcal{A}\mathcal{B}}\check{\chi}_{\mathcal{B}}\;, \chi_\mathcal{A}{}^{\mathcal{B}i}=\varepsilon_{\mathcal{A}\mathcal{C}}\chi^{\mathcal{B}\mathcal{C};i} +\frac{1}{2}\delta_{\mathcal{A}}^\mathcal{B}\varepsilon^{ij}\check{\chi}_j\nonumber \\
&D^{ij}{}_{kl}=\varepsilon^{ij}\varepsilon_{kl}\check{D}, D^{ij}{}_{\mathcal{A}\mathcal{B}}=\varepsilon^{ij}\varepsilon_{\mathcal{A}\mathcal{B}}\check{D},D^{\mathcal{A}\mathcal{B}}{}_{kl}=\varepsilon^{\mathcal{A}\mathcal{B}}\varepsilon_{kl}\check{D}, D^{\mathcal{A}\mathcal{B}}{}_{\mathcal{C}\mathcal{D}}=\varepsilon^{\mathcal{A}\mathcal{B}}\varepsilon_{\mathcal{C}\mathcal{D}}\check{D} \nonumber\\
&D^{i\mathcal{A}}{}_{kl}=\check{D}^{i\mathcal{A}}\varepsilon_{kl}, D^{i\mathcal{A}}{}_{\mathcal{B}\mathcal{C}}=\check{D}^{i\mathcal{A}}\varepsilon_{\mathcal{B}\mathcal{C}}\;, D^{i\mathcal{A}}{}_{j\mathcal{B}}=\varepsilon_{jk}\varepsilon_{\mathcal{B}\mathcal{C}}D^{ik;\mathcal{A}\mathcal{C}}-\frac{1}{2}\delta^i_j\delta^\mathcal{A}_\mathcal{B}\check{D}\nonumber \\
&\phi_{ij}=i\varepsilon_{ij}\phi\;, \phi_{\mathcal{A}\mathcal{B}}=i\varepsilon_{\mathcal{A}\mathcal{B}}\phi\;, \phi_{i\mathcal{A}}=\phi_{i\mathcal{A}} 
\end{align}}
The reality property of some of the decomposed fields is induced from the reality property of the original fields and is given as below:
\begin{align}\label{reality_prop}
    &(V_{\mu}^{ij})^*\equiv V_{\mu ij}=\varepsilon_{ik}\varepsilon_{jl}V_{\mu}^{kl}\;, (V_{\mu}^{\mathcal{A}\mathcal{B}})^*\equiv V_{\mu\mathcal{A}\mathcal{B}}=\varepsilon_{\mathcal{A}\mathcal{C}}\varepsilon_{\mathcal{B}\mathcal{D}}V_{\mu}^{\mathcal{C}\mathcal{D}}\;, (Y_{a}^{i\mathcal{A}})^*\equiv Y_{a i\mathcal{A}} =-\varepsilon_{ij}\varepsilon_{\mathcal{A}\mathcal{B}}Y_{a}^{j\mathcal{B}}\nonumber \\
    &\check{X}_{a}^{*}=\check{X}_{a}\;, (X_{a}^{i\mathcal{A}})^*\equiv X_{a i\mathcal{A}}=\varepsilon_{ij}\varepsilon_{\mathcal{A}\mathcal{B}}X_{a}^{j\mathcal{B}}\nonumber \\
    & \check{D}^*=\check{D}\;, (\check{D}^{i\mathcal{A}})^*\equiv \check{D}_{i\mathcal{A}}=-\varepsilon_{ij}\varepsilon_{\mathcal{A}\mathcal{B}}\check{D}^{j\mathcal{B}}\;, (D^{ij;\mathcal{A}\mathcal{B}})^*\equiv D_{ij;\mathcal{A}\mathcal{B}}=\varepsilon_{ik}\varepsilon_{jl}\varepsilon_{\mathcal{A}\mathcal{C}}\varepsilon_{\mathcal{B}\mathcal{D}}D^{kl;\mathcal{C}\mathcal{D}}\nonumber \\
    & \phi^*=\phi\;, (\phi_{i\mathcal{A}})^*\equiv \phi^{i\mathcal{A}}=\varepsilon^{ij}\varepsilon^{\mathcal{A}\mathcal{B}}\phi_{j\mathcal{B}}
\end{align}
\subsection{Five dimensions}\label{5dirrepsbreaking}
\begin{table}[h!]
	\centering
	\centering
	\begin{tabular}{ |p{4cm}|p{6cm}|}
		\hline
		$USp(4)$ Irreps & $SU(2)\times SU(2)$\\
			\hline
		\multicolumn{2}{|c|}{Fields of the old dilaton Weyl Multiplet}\\
		\hline
$e_\mu^a(\bf{1})$&$e_\mu^a(\bf{1,1})$\\
  $\psi_\mu^I(\bf{4})$ & $\psi_\mu^i(\bf{2,1}),\psi_\mu^{\mathcal{A}}$\\
  $V_\mu^{IJ}(\bf{10}) $ & $V_\mu^{ij}(\bf{3,1})$$,V_\mu^{\mathcal{AB}}(\bf{1,3})$$, Y_a^{i\mathcal{A}}(\bf{2,2}) $\\ 
  $C_\mu (\bf{1}) $ & $C_\mu(\bf{1,1}) $\\
		$T_{ab}^{IJ}$  $(\bf{5})$& $\check{T}_{ab}$ $(\bf{1,1})$,$T_{ab}^{i\mathcal{A}}$  $(\bf{2,2})$\\
		$\chi^{K,IJ}$  $(\bf{16})$&$\check{\chi}^i$ $(\bf{2,1})$, $\check{\chi}^\mathcal{A}$ $(\bf{1,2})$, $\chi^{ij;\mathcal{A}}$ $(\bf{3,2})$, $\chi{}^{i;\mathcal{A}\mathcal{B}}$ $(\bf{2,3})$\\
		$D^{IJ,KL}$ $ (\bf{14})$&$\check{D}$ $(\bf{1,1})$, $\check{D}^{i\mathcal{A}}$ $(\bf{2,2})$, $D^{ij,\mathcal{A}\mathcal{B}}$ $(\bf{3,3})$\\
		\hline
		\multicolumn{2}{|c|}{Fields of the vector Multiplet}\\
		\hline
$A_{\mu }$ $(\bf{1})$& $A_{\mu }$ $(\bf{1,1})$\\
			$\psi^I$ $(\bf{4})$& $\psi^i$ $(\bf{2,1})$, $\psi^\mathcal{A}$ $(\bf{1,2})$\\
			$\sigma^{ij}$ $(\bf{5})$ & $\sigma$ $(\bf{1,1})$, $\sigma^{i\mathcal{A}}$ $(\bf{2,2})$\\
			\hline
	\end{tabular}
	\caption{Decomposition of $USp(4)$ irreps to $SU(2)\times SU(2)$ in five dimensions}
	\label{5dbreak}	
\end{table}
 In table \ref{5dbreak}, we give the details of how the $USp(4)$ irreducible representations carried by the fields of the five dimensional $N=2$ old dilaton Weyl and vector multiplets decompose into $SU(2)\times SU(2) $ irreducible representations. The relations between the fields appearing on the L.H.S of the table and the R.H.S of the table are as given below. We denote some of the components of the fields of the old dilaton Weyl multiplet on the R.H.S of the table, which becomes dependent in the new dilaton Weyl multiplet, with an inverted hat notation.
\allowdisplaybreaks{
\begin{align}\label{2SU(2)}
&V_{\mu}^{ij}=V_{\mu}^{ij}\;, V_{\mu}^{\mathcal{A}\mathcal{B}}=V_{\mu}^{\mathcal{A}\mathcal{B}}\;, V_{a}^{i\mathcal{A}}-u(\psi_a)^{i\mathcal{A}}=Y_{a}^{i\mathcal{A}}\nonumber \\
&T^{ij}_{ab}=\check{T}_{ab}\epsilon^{ij},
 T^{\mathcal{A}\mathcal{B}}=\check{T}_{ab}\epsilon^{\mathcal{A}\mathcal{B}},
 T_{ab}^{i \mathcal{A}}=T_{ab}^{i \mathcal{A}}\nonumber\\
&\chi^{k,ij}=\epsilon^{ij}\check{\chi}^k,
\chi^{\mathcal{A},ij}=\epsilon^{ij}\check{\chi}^{\mathcal{A}},
\chi^{k,\mathcal{A}\mathcal{B}}=\epsilon^{\mathcal{A}\mathcal{B}}\check{\chi}^k,
\chi^{\mathcal{C},\mathcal{A}\mathcal{B}}=\epsilon^{\mathcal{A}\mathcal{B}}\check{\chi}^{\mathcal{C}}\nonumber\\
&\chi^{i,j\mathcal{A}}=\chi^{ij;\mathcal{A}}-\frac{1}{2}\epsilon^{ij}\check{\chi}^{\mathcal{A}},
\chi^{\mathcal{A},\mathcal{B} i}=\chi^{i;\mathcal{A} \mathcal{B}}-\frac{1}{2}\epsilon^{\mathcal{A}\mathcal{B}}\check{\chi}^{i}\nonumber\\
&D^{ij,kl}=\epsilon^{ij}\epsilon^{kl}\check{D},
D^{ij,\mathcal{A}\mathcal{B}}=\epsilon^{ij}\epsilon^{\mathcal{A}\mathcal{B}}\check{D},
D^{\mathcal{A}\mathcal{B},\mathcal{C}\mathcal{D}}=\epsilon^{\mathcal{A}\mathcal{B}}\epsilon^{\mathcal{C}\mathcal{D}}\check{D},\nonumber\\
&D^{i\mathcal{A},kl}=\epsilon^{kl}\check{D}^{i\mathcal{A}},
D^{i\mathcal{A},\mathcal{B}\mathcal{C}}=\epsilon^{\mathcal{B}\mathcal{C}}\check{D}^{i\mathcal{A}}\nonumber\\
&D^{i\mathcal{A},j\mathcal{B}}=D^{ij;\mathcal{A}\mathcal{B}}+\frac{1}{2}\epsilon^{ij}\epsilon^{\mathcal{A}\mathcal{B}}\check{D} \nonumber\\
&\sigma^{ij}=\sigma \varepsilon^{ij}\;,\sigma^{\mathcal{AB}}= \sigma \varepsilon^{\mathcal{AB}}\;, \sigma^{i\AAA}=\sigma^{i\AAA} 
\end{align}
}

\subsection{Six Dimensions}\label{6dirrepsbreaking}

\begin{table}[h!]
	\centering
	\centering
	\begin{tabular}{ |p{4cm}|p{6cm}|}
		\hline
		$USp(4)$ Irreps & $SU(2)\times SU(2)$\\
			\hline
		\multicolumn{2}{|c|}{Fields of the standard Weyl Multiplet}\\
		\hline
  $e_\mu^a(\bf{1})$&$e_\mu^a(\bf{1,1})$\\
  $\psi_\mu^I(\bf{4})$ & $\psi_\mu^i(\bf{2,1}),\psi_\mu^{\mathcal{A}}(\bf{1,2)}$\\
  $V_\mu^{IJ}(\bf{10}) $ & $V_\mu^{ij}(\bf{3,1})$$,V_\mu^{\mathcal{AB}}(\bf{1,3})$$, E_a^{i\mathcal{A}}(\bf{2,2}) $\\ 
		$T_{abc}^{IJ}$  $(\bf{5})$& $T_{abc}$ $(\bf{1,1})$,$T_{abc}^{i,\mathcal{A}}$  $(\bf{2,2})$\\
		$\chi_K{}^{IJ}$  $(\bf{16})$&$\chi_i$ $(\bf{2,1})$, $\chi_\mathcal{A}$ $(\bf{1,2})$, $\chi^{ij;\mathcal{A}}$ $(\bf{3,2})$, $\chi{}^{\mathcal{A}\mathcal{B};i}$ $(\bf{2,3})$\\
		$D^{IJ,KL}$ $ (\bf{14})$&$D$ $(\bf{1,1})$, $D^{i\mathcal{A}}$ $(\bf{2,2})$, $D^{(ij),(\mathcal{A}\mathcal{B})}$ $(\bf{3,3})$\\
		\hline
		\multicolumn{2}{|c|}{Fields of the tensor Multiplet}\\
		\hline
$B_{\mu \nu}$ $(\bf{1})$& $B_{\mu \nu}$ $(\bf{1})$\\
			$\psi^I$ $(\bf{4})$& $\psi^i$ $(\bf{2,1})$, $\psi^\mathcal{A}$ $(\bf{1,2})$\\
			$\phi^{ij}$ $(\bf{5})$ & $\phi$ $(\bf{1,1})$, $\phi^{i\mathcal{A}}$ $(\bf{2,2})$\\
			\hline
	\end{tabular}
	\caption{Decomposition of$ USp (4)$ irreps to $ SU(2) \times SU(2)$ in six dimensions}
	\label{6dbreak}	
\end{table}
 In table \ref{6dbreak}, we give the details of how the $USp(4)$ irreducible representations carried by the fields in the six dimensional $(2,0)$ standard Weyl multiplet and tensor multiplet decompose into $SU(2)\times SU(2) $ irreducible representations. The relations between the fields appearing on the L.H.S of the table and the R.H.S of the table are as given below. We denote some of the components of the fields of the standard Weyl multiplet on the R.H.S of the table, which becomes dependent in the dilaton Weyl multiplet, with an inverted hat notation.
\allowdisplaybreaks{
\begin{align}\label{2SU(2)}
&V_{\mu}^{ij}=V_{\mu}^{ij}\;, V_{\mu}^{\mathcal{A}\mathcal{B}}=V_{\mu}^{\mathcal{A}\mathcal{B}}\;, V_{a}^{i\mathcal{A}}-u(\psi_a)^{i\mathcal{A}}=Y_{a}^{i\mathcal{A}}\nonumber \\
&T^{ij}_{ab}=\check{T}_{ab}\epsilon^{ij},
 T^{\mathcal{A}\mathcal{B}}=\check{T}_{ab}\epsilon^{\mathcal{A}\mathcal{B}},
 T_{ab}^{i \mathcal{A}}=T_{ab}^{i \mathcal{A}}\nonumber\\
&\chi^{k,ij}=\epsilon^{ij}\check{\chi}^k,
\chi^{\mathcal{A},ij}=\epsilon^{ij}\check{\chi}^{\mathcal{A}},
\chi^{k,\mathcal{A}\mathcal{B}}=\epsilon^{\mathcal{A}\mathcal{B}}\check{\chi}^k,
\chi^{\mathcal{C},\mathcal{A}\mathcal{B}}=\epsilon^{\mathcal{A}\mathcal{B}}\check{\chi}^{\mathcal{C}}\nonumber\\
&\chi^{i,j\mathcal{A}}=\chi^{ij;\mathcal{A}}-\frac{1}{2}\epsilon^{ij}\check{\chi}^{\mathcal{A}},
\chi^{\mathcal{A},\mathcal{B} i}=\chi^{i;\mathcal{A} \mathcal{B}}-\frac{1}{2}\epsilon^{\mathcal{A}\mathcal{B}}\check{\chi}^{i}\nonumber\\
&D^{ij,kl}=\epsilon^{ij}\epsilon^{kl}\check{D},
D^{ij,\mathcal{A}\mathcal{B}}=\epsilon^{ij}\epsilon^{\mathcal{A}\mathcal{B}}\check{D},
D^{\mathcal{A}\mathcal{B},\mathcal{C}\mathcal{D}}=\epsilon^{\mathcal{A}\mathcal{B}}\epsilon^{\mathcal{C}\mathcal{D}}\check{D},\nonumber\\
&D^{i\mathcal{A},kl}=\epsilon^{kl}\check{D}^{i\mathcal{A}},
D^{i\mathcal{A},\mathcal{B}\mathcal{C}}=\epsilon^{\mathcal{B}\mathcal{C}}\check{D}^{i\mathcal{A}}\nonumber\\
&D^{i\mathcal{A},j\mathcal{B}}=D^{ij;\mathcal{A}\mathcal{B}}+\frac{1}{2}\epsilon^{ij}\epsilon^{\mathcal{A}\mathcal{B}}\check{D} \nonumber\\
&\phi^{ij}=\phi \varepsilon^{ij}\;,\phi^{\mathcal{AB}}= \phi \varepsilon^{\mathcal{AB}}\;, \phi^{i\AAA}=\phi^{i\AAA} 
\end{align}
}
			\bibliography{references}
			\bibliographystyle{jhep}

\end{document}